\documentclass[useAMS,usenatbib]{mn2e}
\pdfoutput=1
\usepackage{times}
\usepackage{amsmath}    
\usepackage{graphicx}   
\usepackage{verbatim}   
\usepackage{color}      
\usepackage{subfigure}  
\usepackage{amssymb}
\usepackage{cite}
\usepackage{setspace}
\usepackage{gensymb}
\usepackage{hyperref}
\usepackage{array}

\newcommand{\vect}{\mathbf}

\newcommand{\threej}[6]{\left(\begin{array}{ccc} #1 & #2 & #3 \\
                                                 #4 & #5 & #6
                        \end{array}\right)}
\newcommand{\be}{\begin{equation}}
\newcommand{\ee}{\end{equation}}
\newcommand{\ba}{\begin{eqnarray}}
\newcommand{\ea}{\end{eqnarray}}
\newcommand{\bnum}{\begin{enumerate}}
\newcommand{\enum}{\end{enumerate}}

\newcommand{\mnras}{MNRAS}
\newcommand{\aap}{A\&A}

\newcommand{\apj}{ApJ}

\newcommand{\apjs}{APJS}

\newcommand{\jcap}{Journal of Cosmology and Astroparticle Physics}
\newcommand{\prd}{Phys. Rev. D}

\newcommand{\length}{0.23}

\title[A new map-making algorithm for CMB polarisation experiments]{A new map-making algorithm for CMB polarisation experiments}
\author[Wallis et al.]{\parbox[t]{\textwidth}{Christopher G. R. Wallis$^1$\thanks{E-mail:
cwallis@jb.man.ac.uk}, A. Bonaldi$^1$, Michael L. Brown$^1$ and Richard A. Battye$^1$}\vspace*{6pt}\\
$^1$Jodrell Bank Centre for Astrophysics, School of Physics and Astronomy, The University of Manchester, Manchester M13 9PL}
\begin{document}

\date{Accepted 2015 XXXXX XX. Received 2015 XXXXX XX; in original form 2015 XXXXX XX}

\pagerange{\pageref{firstpage}--\pageref{lastpage}} \pubyear{2015}

\maketitle

\label{firstpage}

\begin{abstract}
With the temperature power spectrum of the cosmic microwave background
(CMB) at least four orders of magnitude larger than the $B$-mode
polarisation power spectrum, any instrumental imperfections that
couple temperature to polarisation must be carefully controlled and/or
removed. Here we present two new map-making algorithms that can create
polarisation maps that are clean of temperature-to-polarisation
leakage systematics due to differential gain and pointing between a
detector pair. Where a half wave plate is used, we show that the
spin-2 systematic due to differential ellipticity can also {\color{black} be removed}
using our algorithms. The algorithms require no prior knowledge of
the imperfections or temperature sky to remove the temperature
leakage. Instead, they calculate the systematic and polarisation maps
in one step directly from the time ordered data (TOD). The first
algorithm is designed to work with scan strategies that have a good
range of crossing angles for each map pixel and the second for scan
strategies that have a limited range of crossing angles. The first
algorithm can also be used to identify if systematic errors that have
a particular spin are present in a TOD. We demonstrate the use of both
algorithms and the ability to identify systematics with simulations of
TOD with realistic scan strategies and instrumental noise.
\end{abstract}


\section{Introduction}\label{sec:intro}

The CMB contains an incredible wealth of cosmological information. The
properties of the Universe can be probed at a number of different
epochs using the CMB. The very early universe can be probed through
the CMB's constraints on inflation parameters
\citep{2014A&A...571A..22P}. Physics before last scattering is
imprinted on the CMB as baryon acoustic oscillations, these
oscillations have been mapped to exquisite detail using both the
temperature power spectrum
\citep{2014A&A...571A...1P,2013ApJ...779...86S,2014JCAP...04..014D}
and the $E$-mode polarisation power spectrum {\color{black} 
\citep{2014arXiv1411.1042C,2009ApJ...705..978B,2015arXiv150201582P}}. The large-scale
structure of the universe can also be probed via gravitational lensing
of the CMB. This effect has been measured using high resolution
temperature maps of the CMB
\citep{2014arXiv1412.7521B,2014JCAP...04..014D,2014A&A...571A..17P}.

The CMB $B$-mode polarisation power spectrum contains additional
information on two of these epochs. $B$-mode polarisation on large
angular scales provides us with the best insight into inflation by
placing a direct constraint on the tensor-to-scalar ratio. Tentative
measurements have been made in this area by BICEP2
\citep{2014arXiv1403.4302B}. {\color{black} However,
Galactic foreground emission from polarised dust has been
shown to be responsible for some and possibly all of the signal detected
\citep{2014arXiv1409.5738P,2015PhRvL.114j1301B}.}
 The small scale $B$-mode power spectrum is a result of
gravitational lensing of the larger $E$-mode power spectrum, which has
in recent years been detected by a number of experiments
\citep{2014ApJ...794..171T,2013PhRvL.111n1301H}.

As the $B$-mode power spectrum is at least four orders of magnitude
below the temperature power spectrum any coupling between the two
signals due to instrumental imperfections must be carefully controlled
and/or removed. Approaches used in the literature to ensure the
validity of a polarisation map broadly fall within two categories. The
first of these relies on detailed simulations of the instrumental set
up and uses knowledge of the temperature sky to simulate the effects
of any imperfections on the recovered $B$-mode power spectrum. This
was done very successfully by the POLARBEAR collaboration in their
detection of the lensing $B$-modes \citep{2014ApJ...794..171T}. The
second category involves calculating the coupling by fitting the
parameters describing the imperfections, and then subtracting this
coupling using CMB temperature measurements. This technique was shown
to be effective in the analysis of BICEP2
(\citealt{2014arXiv1403.4302B}, see their figure 5). However, there is
a question as to whether this de-projection technique would work as
effectively with a more complex scan strategy. In addition, the
fitting procedure employed also removes some polarisation signal. This
results in a leakage of $E$-modes to $B$-modes which must be simulated
and removed in the power spectrum estimation
\citep{2015arXiv150200608B}. Here we present alternative novel
algorithms to identify and remove some of the systematics that are
problematic in CMB polarisation experiments. A key feature of our
approach is that it does not require {\it any} prior knowledge of the
telescope or CMB temperature field. In addition, since there is no
fitting involved, our techniques do not result in any leakage of
$E$-modes to $B$-modes.

\citet{2014MNRAS.442.1963W} suggested a map-making algorithm to remove
systematics for experiments where there is no half-wave-plate
(HWP). The method consists of two stages; first systematics of a
different spin to those we want to measure are removed (spin-0 for
temperature and spin-2 for polarisation), then a second cleaning
procedure is required to remove systematics of the same
spin. \citet{2014MNRAS.442.1963W} concentrated on beam
systematics. Consequently, the potential source of spin-2 systematics
that could couple temperature to polarisation that they considered was
the second azimuthal mode of the temperature beam. To remove this they
required knowledge of the beam to correctly predict this leakage from
a temperature map. This method is similar to that used by
\citet{2014arXiv1403.4302B} to remove temperature to polarisation
leakage from differential ellipticity.

One potential complication that the approaches just described suffer
from is the requirement to correctly characterise the ellipticity of
the beam and hence predict the resulting leakage to
polarisation. Conversely, the novel approaches that we present here
require minimal knowledge of the nature of the leakage.  The methods
are appropriate for differencing experiments that use a stepped or
rotating HWP. One of the algorithms can be used even if a HWP is not
present. However, in this case only systematics of a different spin to
polarisation can be removed. Where a HWP is present, it can be used to
disentangle the spin-2 leakage from temperature to polarisation due
the ellipticity of the beam and the spin-2 polarisation signal. The
two methods differ in the scan strategies to which they can be
applied. One algorithm is suited to scan strategies where each map
pixel is seen at a range of telescope orientations, for example the
proposed EPIC scan strategy \citep{2009arXiv0906.1188B}. The other is
suitable for experiments where the range of orientation angles for
each pixel is limited, for example the LSPE scan strategy
\citep{2012SPIE.8446E..7AA}. There is no reason why the two methods
cannot be used on different portions of the same map. For example the
{\it Planck} scan strategy \citep{2014A&A...571A...1P} results in 
good orientation coverage at the ecliptic poles where the first method
would be most suited and a limited range at the ecliptic plane where
the second method would be more appropriate.

We demonstrate that our techniques can also be used to remove
differential gain and pointing even in the absence of a HWP with a
suitable scan strategy. This does leave coupling caused by
differential ellipticity as without a HWP this coupling is
irreducible. In this case we advocate the previous methods of
\citet{2014MNRAS.442.1963W} and \citet{2014arXiv1403.4302B} to remove
this leakage.

Our techniques for removing systematics involve using a model for the
spin of the systematics, which is employed during map-making. This
provides us with $Q$ and $U$ maps that are free of the systematics
included in the model, but also maps of the systematics
themselves. We also demonstrate that our approach can be a useful
method for identifying if a systematic is present in an experiment, or
not.

The paper is organised as follow. Section \ref{sec:mapmaking_hwp}
describes the analytical framework for the algorithms to remove
systematics. Then in Section \ref{sec:sim_test} we demonstrate the use
of the algorithms on simulations, using realistic scan strategies and
time ordered data (TOD) which include instrumental noise. Section
\ref{sec:identify_sys} explains how the algorithm can be used to find
systematics and demonstrates the technique on a simulated TOD. Finally
in Section \ref{sec:discuss} we summarise our work.

\section{Map-making algorithms}\label{sec:mapmaking_hwp}

Our objective is to create maps free of systematic error due to the
imperfections in the instrumentation of an experiment. We assume the
HWP is ideal and situated at the end of the optical system (in
emission). The effect of the HWP is to simply rotate the angle of the
polarisation sensitivity of the beam (see,
e.g.~\citealt{2009MNRAS.397..634B}), leaving the shape of the
polarisation intensity and temperature beams unchanged.  {\color{black}
  The assumption of an ideal HWP is obviously not entirely
  realistic. However, in practice a HWP would only ever be included in
  an experiment if the systematic effects that they introduce are
  smaller than the effects that they are designed to
  mitigate. Relaxing the assumption of HWP ideality is something we
  leave to further work.} We consider an experiment where a detector
pair is used to measure temperature and polarisation. {\color{black}
  Each detector is sensitive to orthogonal polarisation directions.}
The two signals $d_1$ and $d_2$ are summed and differenced:
\ba
S^{\rm add} &=& \frac{1}{2}(d_1 + d_2),\\
S^{\rm dif} &=& \frac{1}{2}(d_1 - d_2).\label{eq:dif_signal}
\ea
In an ideal experiment $S^{\rm add}$ would correspond to the
temperature of the pixel and $S^{\rm dif}$ the rotated polarisation,
the only effect of the beam would be to isotropically smooth the
temperature and polarisation fields. 

We will concentrate on recovering
the polarisation of the pixel and therefore, we drop the superscript in
equation \eqref{eq:dif_signal} at
this stage. Therefore, the differenced signal, $S$, will be the
rotated polarisation of the pixel plus any systematics, the most
serious of which will couple temperature to polarisation. Some common
systematics include differential gain, differential pointing and
differential ellipticity of the detector pair. These systematics
transform as spin-0, spin-1 and spin-2 respectively with telescope
orientation $\psi_{\rm t}$. For a demonstration of the leakage angular
dependence see e.g.~fig 2 of \citet{2008PhRvD..77h3003S} where the
authors depict the monopole, dipole and quadrupole nature of the
different systematics. {\color{black}The differential gain is spin-0 as it
is simply a scaled temperature map. The differential pointing is spin-1
as the signal is a difference of the temperature map at two close points
in space. The leakage, to first order, is therefore proportional to the 
derivative of the beam smoothed temperature field.
 Differencing two elliptical Gaussians results in a quadrupole pattern. This quadrupole pattern
is then convolved with the sky to create a spin-2 systematic effect.
The systematic errors must be constant for this map-making algorithm
to be able to remove them. If the systematics change with time, a more
adaptive algorithm would need to be developed.}

With the HWP at the end of the optical system there is no dependence
of these systematics on the orientation of the HWP $\psi_{\rm h}$. The
detected differenced signal $S$ is therefore,
\ba
S(\psi_{\rm h}, \psi_{\rm t}) &=& \Re\left[P e^{i2(\psi_{\rm t} + 2\psi_{\rm h})} + G + M e^{i\psi_{\rm t}} + E e^{i2\psi_{\rm t}}\right],\\
S(\psi_{\rm r}, \psi_{\rm t}) &=&\Re\left[P e^{i2\psi_{\rm r}} + G + M e^{i\psi_{\rm t}} + E e^{i2\psi_{\rm t}}\right] \label{eq:rot_sys},
\ea
where $P$ is the complex representation of the polarisation of the
pixel in terms of the Stokes parameters, $P=Q+iU$. $G, M$ and $E$ are
the temperature to polarisation leakage due to differential gain,
differential pointing and differential ellipticity respectively, and $\Re$
is the real part operator. The magnitudes and phases of the
systematics are dependent on the nature of the imperfections and the
underlying temperature field. Note however that the magnitudes and
phases are unimportant for this work. Here, only knowledge of the way they
transform with the telescope orientation is required in order to remove the
systematics. In equation \eqref{eq:rot_sys} we have made a coordinate
transformation $\psi_{\rm r} {=} \psi_{\rm t} {+} 2\psi_{\rm h}$. We
do this so that the polarisation and systematics are dependent on
different variables in our space.

The aim of this work is, therefore, to obtain an unbiased estimate of
$P$, given that the detected signal depends on the systematics as well
as polarisation. The two techniques which we present differ only in
the scan strategies that they can be applied to. We first present an
algorithm suitable for a scan strategy where the $\psi_{\rm t}$
coverage of a pixel is extensive. For example, the EPIC
\citep{2009arXiv0906.1188B} strategy is designed to maximise this
coverage. We then present a second method where the $\psi_{\rm t}$
coverage is limited. Balloon borne experiments such as LSPE
\citep{2012SPIE.8446E..7AA} will have limited $\psi_{\rm t}$
coverage. Such experiments often include a rotatable HWP in order to
obtain multiple polarisation crossing angles.

{\color{black} When a rotating or stepped HWP is included in an
  experiment, whatever the scan strategy, the HWP can be used to
  provide enough polarisation angle coverage such that detector
  differencing is not required. As the differencing seems to lead to
  the temperature to polarisation leakage considered in this work, one
  may ask if other techniques, which do not require differencing, could
  be used. If the HWP is continually rotating then certain ``lock-in''
  techniques can be used to isolate the polarisation signal from the
  systematic errors \citep{2007ApJ...665...55W}. However, maintaining
  continuous rotation of the HWP can cause its own wealth of
  systematic errors. We therefore focus of the case of a stepped HWP
  for which ``lock-in'' techniques are not applicable.  

Even with a stepped HWP, the large amount of polarisation angles
provided by the HWP in principle allows one to recover maps of the
Stokes parameters from just one detector. Such a technique is however
more problematic than differencing as the temperature to polarisation
leakage could potentially be much worse. A differencing experiment
allows two detectors, that are located at exactly the same position in
the focal plane (and therefore observe the same point on the sky) to
be used to directly remove the temperature signal (see
equation~\ref{eq:dif_signal}). If a single detector was used to
reconstruct Stokes parameter maps, then the absolute pointing error,
which is typically larger than the differential pointing considered
here, would create different temperature responses between different
observations of a pixel and this would leak temperature fluctuations
to polarisation.

A similar argument holds for ellipticity; by differencing detector
pairs, we are susceptible to the difference in the ellipticity of two
beams which often have very similar beam shapes. By creating
polarisation maps from one detector the total ellipticity would create
different temperature responses when the telescope observes a pixel at
different orientations, leading to a much larger temperature to
polarisation leakage. One problem that detector differencing can
suffer from, and that using one detector avoids, is a constant
differential calibration. However, in this case the benefits of
differencing often outweigh this particular problem. Another benefit
in differencing two detectors, is that correlated noise between the
detectors is removed. Motivated by these considerations we have
adopted a map-making scheme that differences two detectors in a
detector pair.}

\subsection{Map-making algorithm with extensive $\psi_{\rm t}$ coverage} \label{sec:psi_t_ext}

Our experimental model consists of sampling a pixel at a wide range of
orientations of the telescope and HWP. The detected signal, $S^d$, can
be expressed as,
\ba
S^d(\psi_{\rm r}, \psi_{\rm t}) &=& h(\psi_{\rm r}, \psi_{\rm t})S(\psi_{\rm r}, \psi_{\rm t}),~ \text{where}\label{eq:detected_signal}\\
h(\psi_{\rm r}, \psi_{\rm t}) &\equiv& \frac{1}{N_{\rm hits}}\sum_{i{=}1}^{N_{\rm hits}} \delta(\psi_{\rm r} -\psi_{\rm r}^i)\delta(\psi_{\rm t} -\psi_{\rm t}^i)
\ea
is the window function representing the knowledge that we have of the
pixel. One sample, $i$, will contribute one delta function
$\delta(\psi_{\rm r} -\psi_{\rm r}^i)\delta(\psi_{\rm t} -\psi_{\rm
  t}^i)$ to this window. Our aim is to obtain an unbiased estimate of
the polarisation of the pixel given that the systematics are present
and have the functional form outlined in equation
\eqref{eq:rot_sys}. This functional form lends itself to be described
well by a Fourier series. We replace each term in equation
\eqref{eq:detected_signal} with their Fourier series such that,
\begin{eqnarray}
\sum_{n_1 m_{1}} &&\!\!\!\!\!\!\!\!\!\!\!\!\tilde{S}^d_{n_1, m_1} e^{i(n_1\psi_{\rm r} +m_1\psi_{\rm t})}= \nonumber \\
&& \!\!\!\!\!\sum_{\substack{n_2 m_2 \\n_3 m_3}} \tilde{h}_{n_2, m_2} e^{i(n_2\psi_{\rm r} +m_2\psi_{\rm t})} \tilde{S}_{n_3, m_3} e^{i(n_3\psi_{\rm r} +m_3\psi_{\rm t})}.
\ea
Multiplying each side by $\frac{1}{8\pi^2}e^{-i(N\psi_{\rm r}
  +M\psi_{\rm t})}$, integrating over the whole $(\psi_{\rm
  r},\psi_{\rm t})$ space and evaluating the resulting Kronecker
delta function, we find
\ba
\tilde{S}^d_{n_1 m_1} &=& \frac{1}{8\pi^2}\sum_{\substack{n_2 m_2 \\n_3 m_3}} \tilde{h}_{n_2, m_2} \tilde{S}_{n_3, m_3}\nonumber\\
&&\!\!\!\!\!\!\!\!\!\!\times \int_0^{4\pi}\!\!\!\!\!\!d\psi_{\rm r}\int_0^{2\pi} \!\!\!\!\!\!d\psi_{\rm t} \;e^{i[(n_2+n_3-n_1)\psi_{\rm r} + (m_2 + m_3 - m_1)\psi_{\rm t})]},\\
 &=& \sum_{n_3 m_3} \tilde{h}_{n_1{-}n_3, m_1{-}m_3} \tilde{S}_{n_3, m_3}.\label{eq:2d_for}
\ea
%

In principle, an unbiased estimator for the different components of
the signal can now be formed by inverting equation
\eqref{eq:2d_for}. However, this operation is not yet possible for two
reasons. Firstly, we are attempting to invert a matrix infinite in
size. Secondly for any realistic window function\footnote{By realistic
  we specifically mean any window function where $h(\psi_{\rm r},
  \psi_{\rm t})=1$.} the {\color{black} matrix} will be singular. By understanding the
dependence on $\psi_{\rm r}$ and $\psi_{\rm t}$ of $S(\psi_{\rm r},
\psi_{\rm t})$, we can ignore terms in equation \eqref{eq:2d_for}
where $\tilde{S}_{n_3, m_3}{=} 0$, thereby making the operation
invertible and obtaining an unbiased estimate of $\tilde{S}_{n_3,
  m_3}$ from our detected $\tilde{S}^d_{n_3, m_3}$. If we know the
differenced signal contains temperature to polarisation leakage from
differential gain we would include the term $\tilde{S}_{0,0}$. For
differential pointing and ellipticity, we include $\tilde{S}_{0,\pm1}$
and $\tilde{S}_{0,\pm2}$ respectively. In principle we could remove
systematics of any spin by simply including the correct term. The
polarisation of the pixel will be,
\ba
Q &=& 2\Re(\tilde{S}_{2,0}),\\
U &=& 2\Im(\tilde{S}_{2,0}).
\ea
Equation \eqref{eq:2d_for} will only be invertible if there are enough
hits on the pixel at a sufficient variety of crossing angles
$\psi_{\rm t}$ and HWP angles $\psi_{\rm h}$. The more terms we
include in equation \eqref{eq:2d_for} the more observed orientations
will be required.

\subsection{Map-making algorithm with limited $\psi_{\rm t}$ coverage} \label{sec:psi_t_lim}

The second class of experiments that we consider has a limited range
of crossing angles $\psi_{\rm t}$ and obtains polarisation angle
coverage using a stepped HWP. This is similar to the observation
strategy envisaged for the LSPE \citep{2012SPIE.8446E..7AA}. In this
case using Fourier terms to describe the systematics is not a good
choice. Here we describe a formalism that is specifically designed for
a small, but non-zero, range of crossing angles.

We start from the same position as for the case of extensive
$\psi_{\rm t}$ coverage in Section \ref{sec:psi_t_ext}. In an
experiment we have a function describing the detected signal given by
equation \eqref{eq:detected_signal}. However the range of angles
$\psi_{\rm t}$ is small. This restricted range of angles means that
describing the full Fourier mode of each systematic would be
problematic. Instead, we choose to describe the summed effect of the
systematics in terms of Legendre polynomials. Let the $\psi_{\rm t}$
angles range from $\psi_{\rm t}^{\rm min}$ to $\psi_{\rm t}^{\rm
  max}$. We can now define a coordinate that spans this range:
\ba
x  = \frac{2(\psi_{\rm t} - \psi_{\rm t}^{\rm min})}{\psi_{\rm t}^{\rm max} - \psi_{\rm t}^{\rm min}} - 1,
\ea
where $x$ ranges from $-1$ to $1$. With this definition we can now
rewrite equation \eqref{eq:rot_sys} as
\ba
S(\psi_{\rm r}, x) &=&\Re\left[P e^{i2\psi_{\rm r}}\right] + f(x), \label{eq:limit_model}
\ea
where $f(x)$ is a function that describes the combined effects of the
systematic leakage from temperature to polarisation. If the $\psi_{\rm
  t}$ range is small enough then $f(x)$ will be well described by only
a few Legendre polynomials. In Fig. \ref{fig:leg_demo} we show the
effectiveness of the Legendre polynomials to describe a particular
section of a function of the form $g(\psi)=\cos(2\psi + \pi/8)$, where
the $\psi$ range is $\pm0.25$ rads. This range is chosen to
approximate the range of crossing angles seen in typical balloon
experiments. In particular, the maximum $\psi_{\rm t}$ range in any
pixel in the LSPE scan strategy \citep{2012SPIE.8446E..7AA} is
$\approx$0.5 rads. The left panel shows that the amplitudes reduce almost
exponentially with the order of the polynomial. In the centre and
right panels we demonstrate that using only the first 3 Legendre
polynomials, we can reconstruct the systematic to within fractions of
a percent.

\begin{figure*}
\begin{center}

~\\
~\\
\includegraphics[width=0.293\linewidth, trim=0cm 0cm 0cm 0cm, clip=true]{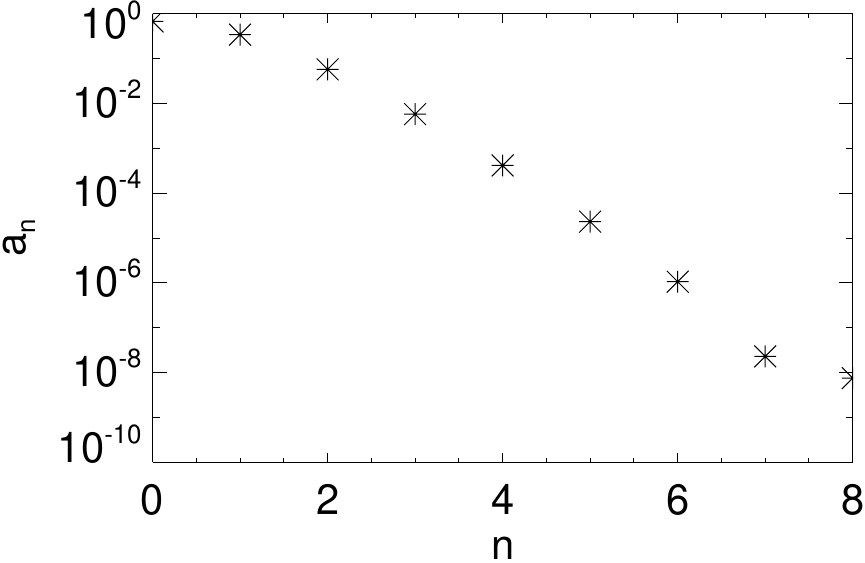}  
\includegraphics[width=0.293\linewidth, trim=0cm 0cm 0cm 0cm, clip=true]{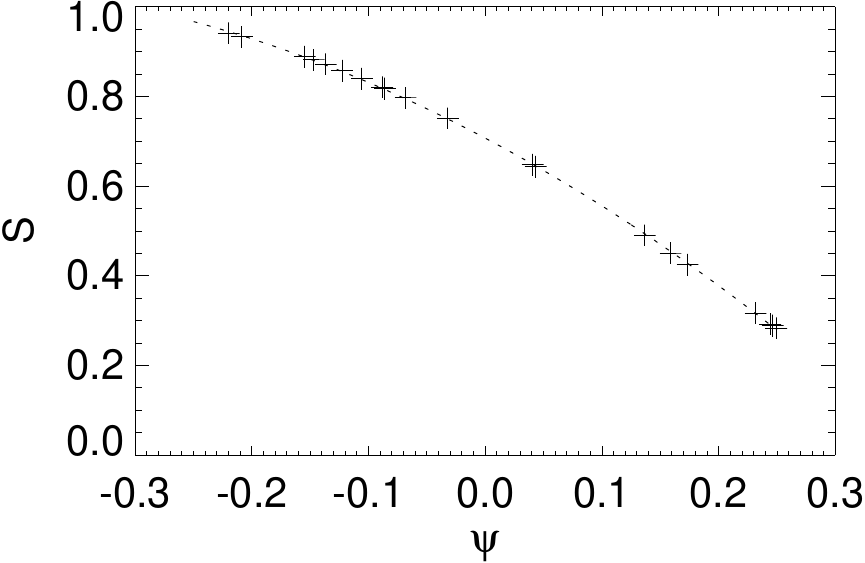}
\includegraphics[width=0.314\linewidth, trim=0cm 0cm 0cm 0cm, clip=true]{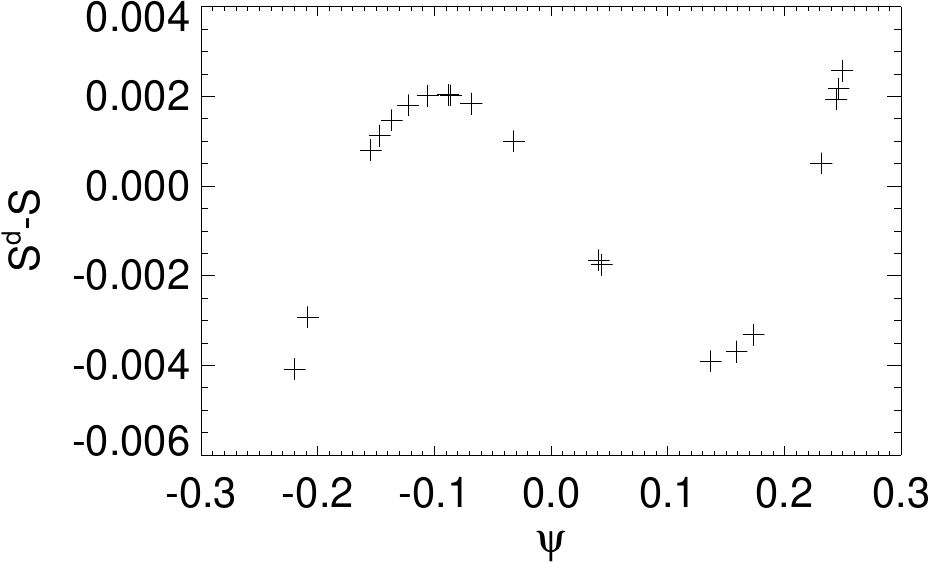}\\

\caption{{\it Left panel:} The amplitudes of Legendre polynomials 
required to describe $g(\psi)=\cos(2\psi + \pi/8)$, where the
$\psi$ range is $\pm0.25$ rads. This range is typical of the $\psi_{\rm t}$ 
range of LSPE. {\it Centre panel:} A demonstration of using the first 3 
Legendre polynomials to describe the systematic $g(\psi)$, the crosses
are the "hits" which were randomly generated with a uniform distribution 
over the full range. The dashed line shows the reconstruction of the 
systematic. {\it Right panel:} The error between the reconstruction and the
input systematic. The functional form of the residual looks very similar
to that of the next polynomial in the series, $P_3(x)$.
\label{fig:leg_demo}}
\end{center}
\end{figure*}

With this motivation we follow similar steps to those in Section
\ref{sec:psi_t_ext} to create an estimate of the polarisation of a
pixel free of systematics. We use the Legendre polynomials to describe
the $x$ dependence of the signal and a Fourier series to describe the
$\psi_{\rm r}$ dependence. To begin we write the problem as a multiple
of the underlying signal and the window function,
\ba
S^d(\psi_{\rm r}, x) &=& h(\psi_{\rm r}, x)S(\psi_{\rm r}, x).\label{eq:dectected_signal_x}
\ea
As before we substitute the functions for their decompositions into a
set of basis functions, where here we have chosen the Legendre
polynomials:
\ba
&&\!\!\!\!\!\!\!\!\!\!\!\!\!\!\sum_{n_1 m_{1}} \tilde{S}^d_{n_1, m_1} e^{in_1\psi_{\rm r}} P_{m_1}(x) = 
\nonumber \\
&&\;\;\;\;\;\;\;\;\sum_{\substack{n_2 m_2 \\n_3 m_3}} \tilde{h}_{n_2, m_2} e^{in_2\psi_{\rm r}} P_{m_2}(x) \tilde{S}_{n_3, m_3} e^{in_3\psi_{\rm r}}P_{m_3}(x)
\ea
Taking the scalar product\footnote{The scalar product we use is,
\ba
\int_0^{4\pi} \!\!\!\!\!\!d\psi_{\rm r}\int_{-1}^{1} \!\!\!\!\!\!dx f(\psi_t,x)g(\psi,x).
\ea}
 of both sides with a basis function leaves us with the triple integral,
\ba
\tilde{S}^d_{n_1, m_1} &=& \frac{2m_1+1}{8\pi}\sum_{\substack{n_2 m_2 \\n_3 m_3}} \tilde{h}_{n_2, m_2}\tilde{S}_{n_3, m_3}\nonumber\\
&& \!\!\!\!\!\!\!\!\!\!\!\!\!\!\!\!\!\!\!\!\!\!\!\!\!\!\!\!\times\int_0^{4\pi} \!\!\!\!\!\!d\psi_{\rm r}\int_{-1}^{1} \!\!\!\!\!\!dx \;e^{i[(n_2+n_3-n_1)\psi_{\rm r}} P_{m_1}(x)P_{m_2}(x)P_{m_3}(x),\\
\tilde{S}^d_{n_1, m_1} &=& (2m_1+1)\sum_{m_2 n_3 m_3} \tilde{h}_{n_1{-}n_3, m_2} \tilde{S}_{n_3, m_3}\nonumber\\
&&\;\;\;\;\;\;\;\;\;\;\;\;\;\;\;\;\;\;\;\;\;\;\;\;\;\;\times\threej{m_1}{m_2}{m_3}{0}{0}{0}^2,\label{eq:2d_forleg}
\ea
where we have used the Wigner 3j symbol, 
\ba
\threej{m_1}{m_2}{m_3}{\ell_1}{\ell_2}{\ell_3}.
\ea
 Once again we can obtain an unbiased estimate of the polarisation by
 calculating this coupling matrix and then inverting it. Explicitly
 the polarisation will be
\ba
Q &=& 2\Re(\tilde{S}_{2,0}),\\
U &=& 2\Im(\tilde{S}_{2,0}).
\ea
Just as in Section \ref{sec:psi_t_ext} where we had to ensure that we
included all the Fourier modes of the systematics, here we will have
to include all of the Legendre polynomials that describe $f(x)$. This
will depend on the underlying systematics and the range of $\psi_{\rm
  t}$ angles seen at each pixel. Unlike in Section \ref{sec:psi_t_ext}
where the term chosen is a direct result of the spin of the systematic
required to be removed, here there is no physical motivation for the
terms to use. We simply require that enough terms are used such that
we obtain a satisfactory fit for the combined result of the
systematics, $f(x)$.

\section{Test on Simulations}\label{sec:sim_test}

To test the map making algorithms described in Section
\ref{sec:mapmaking_hwp} we simulate two common types of experiment:
one satellite-like experiment, having an extensive range of
orientation angles, and one balloon-like experiment where this range
is limited. To this end, we use the EPIC \citep{2009arXiv0906.1188B}
and the LSPE \citep{2012SPIE.8446E..7AA} scanning strategies
respectively. The hit maps of the two scan strategies are shown in
Fig.~\ref{fig:hit_maps}. {\color{black} In this figure, we also plot 
the polarisation angle coverage ($p_2$) for each pixel,
\ba
p_2 = \frac{1}{N_{\rm hits}}\sum_{i=1}^{N_{\rm hits}} [\cos^2(2\psi_t^i) + \sin^2(2\psi_t^i)], \label{eq:pol_cov}
\ea
where $N_{\rm hits}$ is the number of hits that a pixel has received. 
This quantity demonstrates the range of $\psi_t$ angles provided by
the scan strategy. The range of $p_2$ goes from 0 to 1 and the lower
the value, the better the polarisation angle coverage.}

\begin{figure*}
\begin{center}
\begin{tabular}{c c}
\includegraphics[width=0.45\linewidth, angle=180, trim=0cm 1cm 0cm 0cm, clip=true]{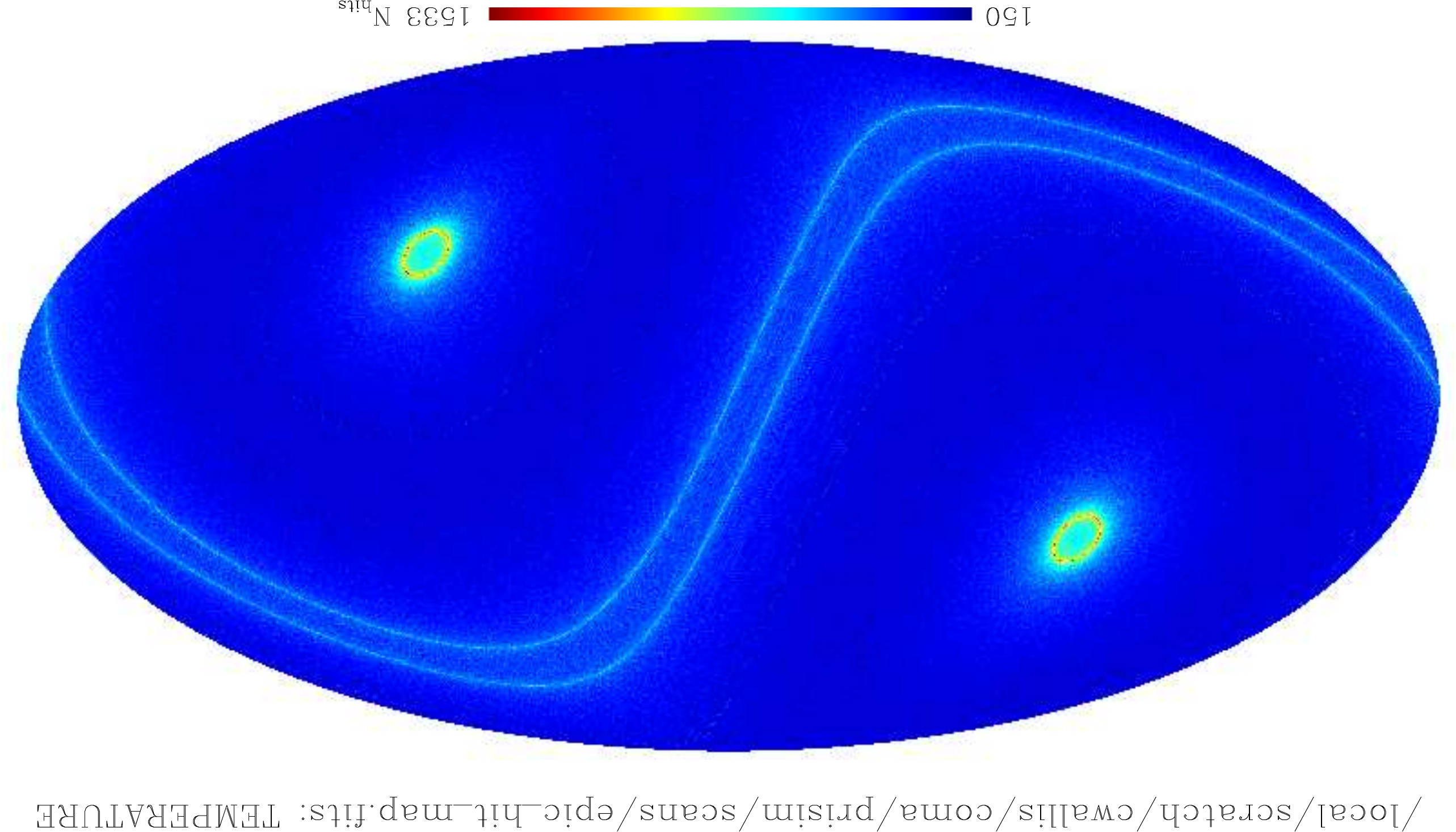} &
\includegraphics[width=0.45\linewidth, angle=180, trim=0cm 1cm 0cm 0cm, clip=true]{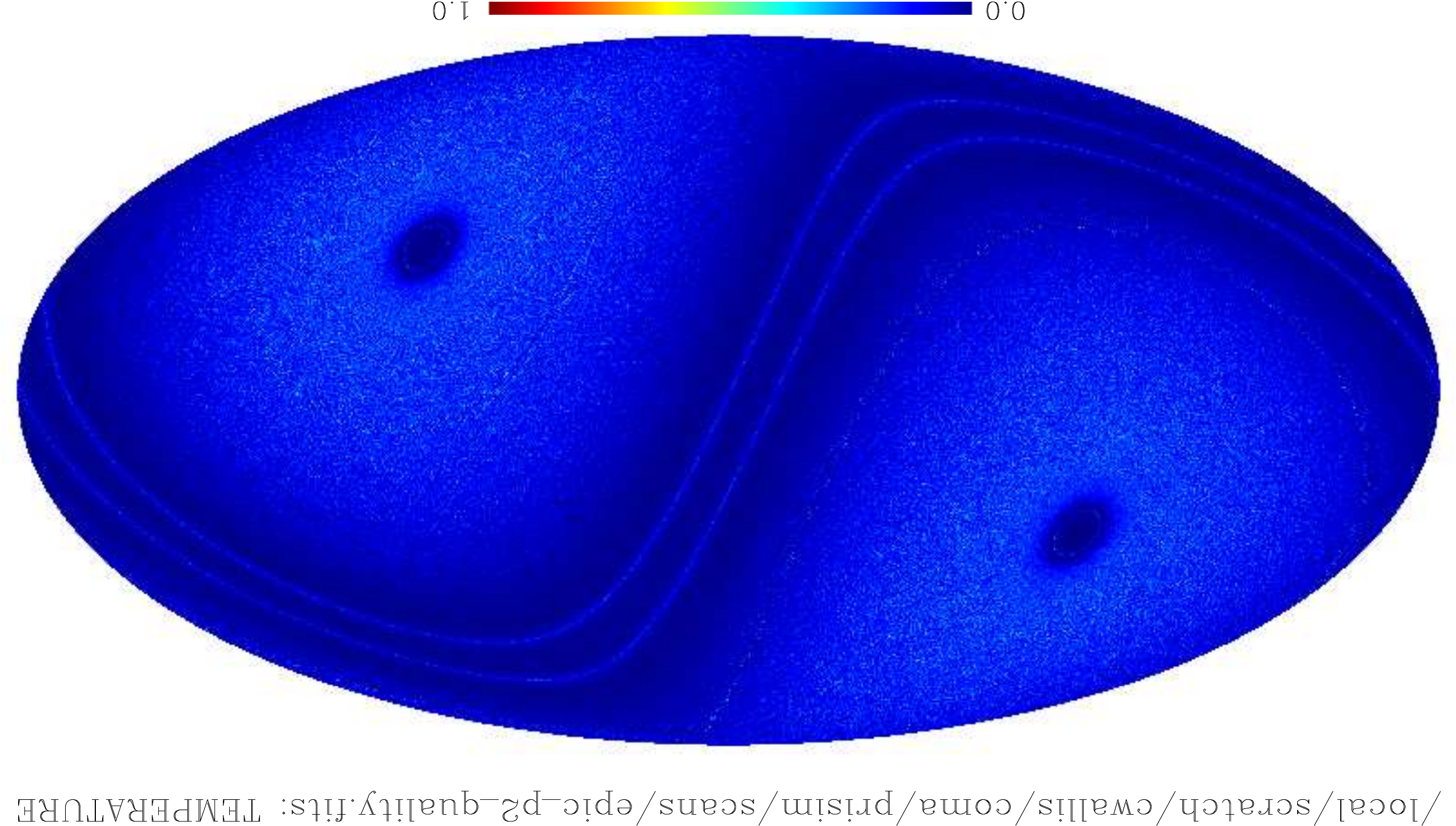}\\
\includegraphics[width=0.45\linewidth, angle=180, trim=0cm 1cm 0cm 0cm, clip=true]{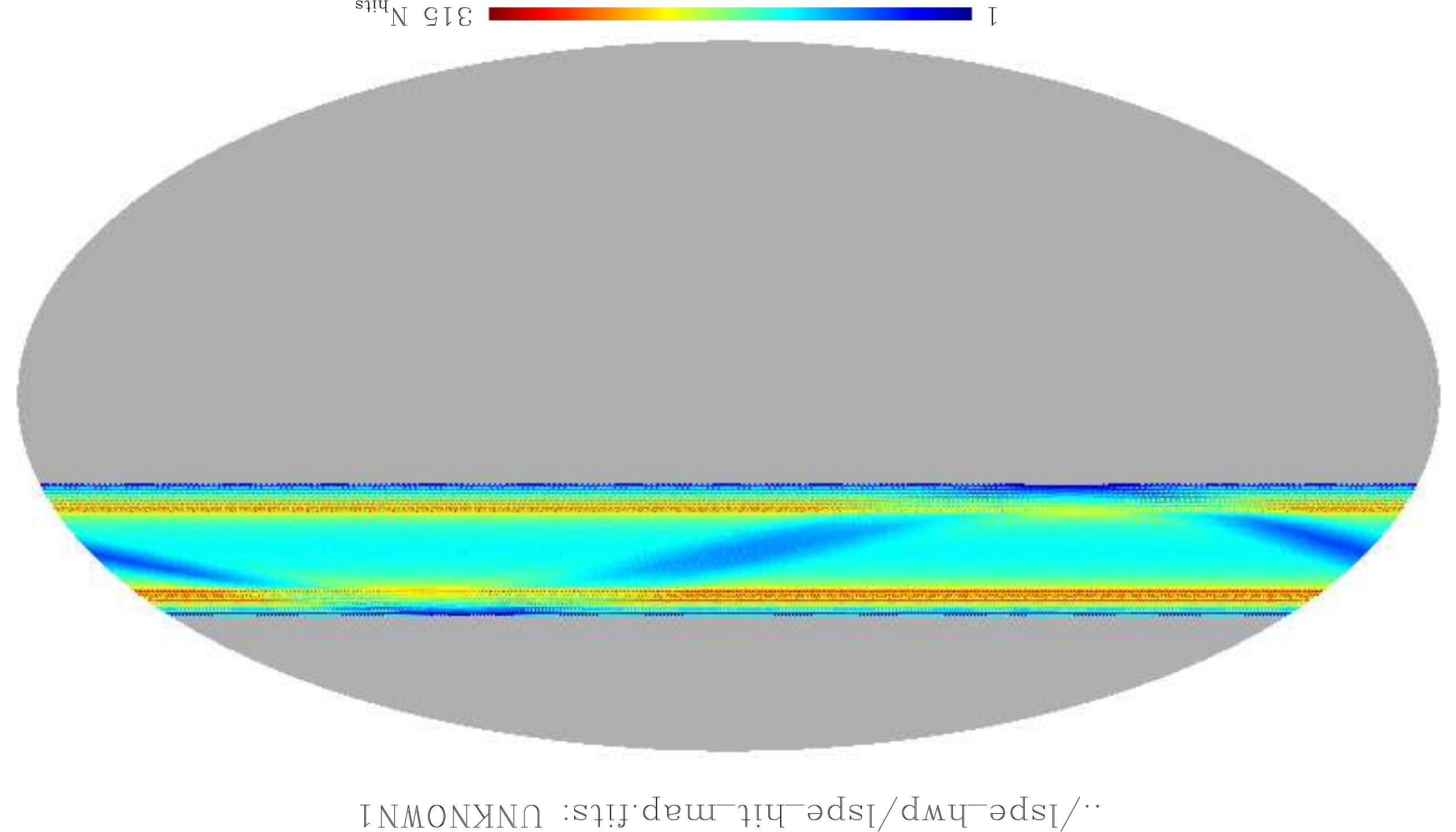} &
\includegraphics[width=0.45\linewidth, angle=180, trim=0cm 1cm 0cm 0cm, clip=true]{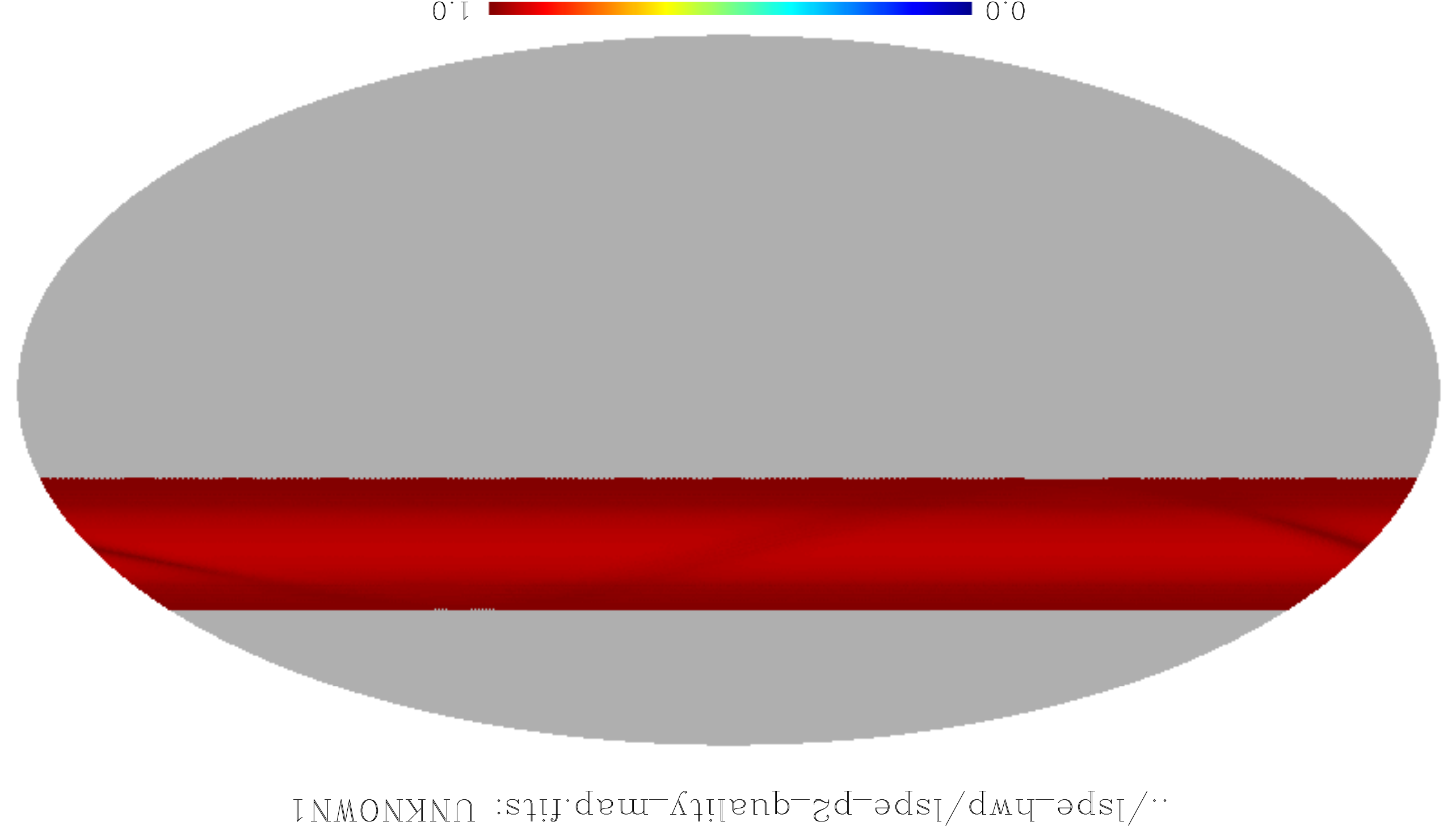}\\
\end{tabular}
\vspace{0.5cm}
\caption{{\color{black}The hit maps (\emph{left}) and polarisation
    coverage (\emph{right}, $p_2$), see equation \eqref{eq:pol_cov},
of the two scan strategies used in the
  simulations. The plots show a full sky Mollweide projection of the sphere.
{\it Upper panels:} the EPIC scan strategy
  \citep{2009arXiv0906.1188B} used to demonstrate the ``extensive
  $\psi_{\rm t}$" range algorithm. The EPIC scan strategy is designed
  to mitigate systematic errors by having many crossing angles for
  each pixel. This projection is in Galactic coordinates. 
{\it Lower panels:} the LSPE scan strategy
  \citep{2012SPIE.8446E..7AA} used to demonstrate the ``limited
  $\psi_{\rm t}$" range algorithm. The LSPE gondola will perform scans
  of constant azimuth, changing the elevation $\sim$daily. There is,
  therefore, a limited range of crossing angles for each pixel. LSPE
  will obtain good polarisation angle coverage using a stepped HWP.
This projection is in Ecliptic coordinates.
}\label{fig:hit_maps}}
\end{center}
\end{figure*}


In all the simulations we use a fiducial power spectrum with a
scalar-to-tensor ratio of 0.1 and lensing $B$-modes are also present.

\subsection{Extensive $\psi_{\rm t}$ coverage algorithm with a satellite-like experiment}\label{sec:epic_sims}

We simulate noisy TODs with systematic errors. The main source of
systematics we will be considering are leakage from temperature
fluctuations to polarisation fluctuations. Therefore it may be
sufficient to only simulate the systematics introduced by
imperfections that couple temperature to polarisation. A TOD element,
$t_j$, is simply the temperature and polarisation response multiplied
by the underlying CMB sky and then integrated,
{\color{black}
\ba 
t_j &=& G\int \vect{du} [B^T_j(\vect{u})T(\vect{u}+\Delta\vect{p}) +
  B^Q_j(\vect{u})Q(\vect{u}+\Delta\vect{p}) \nonumber\\
&& \qquad\qquad\qquad\qquad\qquad\quad+ B^U_j(\vect{u})U(\vect{u}+\Delta\vect{p})], \label{eq:full_tj} 
\ea }
where $X(\vect{u})$ is the sky emission in the Stokes parameter $X$
from the direction pointed to by the unit vector $\vect{u}$.
$B^X_j(\vect{u})$ is the beam response in the direction $\vect{u}$ for
the Stokes parameter $X$ when orientated in the position $j$. 
{\color{black}$G$ is the gain of the detector and $\Delta\vect{p}$
 is the shift in the temperature beam due to the pointing error.} The
position $j$ describes the orientation of the telescope by the
standard Euler angles and the orientation of the HWP. In order to
simulate this correctly one would need to convolve the sky over this 4
dimensional space. For a high resolution experiment this would be
computationally infeasible, especially when one requires many CMB
realisations. We therefore only simulate beam systematics that couple
temperature to polarisation due to differential ellipticity. The
second and third term of the RHS of equation \eqref{eq:full_tj} can be
calculated simply from a polarisation map of the sky smoothed with the
axisymmetric component of the beam. In our current model of the HWP in
CMB experiments, the orientation of the HWP ($\psi_{\rm h}$) has no effect
on the temperature response. Therefore, we only require the
convolution of the beam over the 3 dimensional space
$(\theta, \phi, \psi_{\rm t})$. This approximation can be formally
written as,
{\color{black}
\ba
t_j &=& G\int \vect{du} [B^T_j(\vect{u})T(\vect{u} + \Delta\vect{p})+ A^Q_j(\vect{u})Q(\vect{u}+\Delta\vect{p})\nonumber\\
&& \qquad\qquad\qquad\qquad\qquad\quad + A^U_j(\vect{u})U(\vect{u}+\Delta\vect{p})], \label{eq:simplified_tj}
\ea}where
 $A^X_j(\vect{u})$ is the axisymmetric component of the beam
response. The first term of equation~\eqref{eq:simplified_tj} is
calculated by a fast pixel space convolution code developed in
\citet{2014MNRAS.442.1963W} based on the algorithm described in
\citet{2011ApJS..193....5M}. The code produces the temperature field
convolved with the asymmetric beam as binned in the 3 dimensional
space. In the $\theta$ and $\phi$ space we use a {\sevensize HEALPix}
pixellation \citep{2005ApJ...622..759G}, and in the $\psi_t$ space we
use a linear binning. The convolution code calculates the central
values of the pixels for this 3 dimensional grid. We use $N_{\rm
  side}{=}2048$ for the {\sevensize HEALPix} pixelisation and the
$\psi_{\rm t}$ space is separated into 80 bins. The second and third
terms of equation \eqref{eq:simplified_tj} are calculated using the
{\sevensize SYNFAST} program part of the {\sevensize HEALPix}
package. Each of these codes gives us the central values of the
pixelised space. We therefore, use linear interpolation to calculate
the TOD element for a particular pointing.

We use this set up to simulate the TODs for one detector pair for a
given scan strategy. For the temperature beam, $B^T$, we use an
elliptical Gaussian described by
\ba
B^T(\theta,\phi) = \frac{1}{2\pi q \sigma^2} e^{-\frac{\theta^2}{2\sigma^2}(\cos^2\phi + q^{-1}\sin^2\phi)}. \label{eq:gaus_beam}
\ea
Equation~(\ref{eq:gaus_beam}) describes the beam for detector 1. The other detector has a similar profile
except it is rotated by $\pi/2$ to create a differential ellipticity
between the two detectors. We use $\sigma = 3$ arcmin corresponding to
a FWHM of 7 arcmin and the ellipticity parameter $q = 1.2$.

We include a differential gain between the detectors by simply
multiplying one detector's response by a constant gain factor. We also
simulate a constant differential pointing by simply including an
offset in one of the detector pointings in our simulation.

We use the EPIC scan strategy \citep{2009arXiv0906.1188B} in the
following simulations with and without a stepped HWP, to simulate one
detector pair that suffers from the systematics we consider in this
paper. See Fig.~\ref{fig:hit_maps} for the hit map of the EPIC scan
strategy. We step the HWP by $\pi/8$ every 1hr. For these satellite
simulations we do not include a Galactic mask. This map-making
algorithm works in a very similar way to a binned map and only
requires the TOD data from one pixel to create an estimate of the
Stokes $Q$ and $U$ of a pixel. It will therefore work equal well
regardless of the sky coverage. Here we use the entire sky to make the
power spectrum analysis simple.

\subsubsection{Simulation 1: No HWP included, no noise}\label{sec:sim1}

We simulate a noise-free TOD from a detector pair that suffers from a
differential gain of the two detectors of 1\% and a differential
pointing of 0.1 arcmin which is 1.5\% of the 7 arcmin (FWHM) beam. We do not
include any differential ellipticity because in this simulation we do
not have a HWP. Without a HWP the spin-2 systematic created by
differential ellipticity cannot be distinguished from the spin-2
polarisation signal. Therefore, the technique we present in this paper
cannot remove the systematic. If an experiment needs to remove this
systematic we suggest the methods presented in
\citet{2014arXiv1403.4302B} and
\citet{2014MNRAS.442.1963W}. Fig.~\ref{fig:epic_b_mode} 
shows the recovered $B$-mode power spectrum when
a simple binned map is made from this TOD compared to one where the
algorithm described in Section \ref{sec:psi_t_ext} is used. We
included the terms $\tilde{S}_{0, 0}$ and $\tilde{S}_{0, 1}$ in
equation \eqref{eq:2d_for} to account for the differential gain and
pointing. In Fig.~\ref{fig:epic_b_mode} we can clearly see that the
algorithm has removed the bias on the recovered $B$-mode power
spectrum as a result of the temperature to polarisation leakage.
\begin{figure*}
\begin{center}
\begin{tabular}{c c}
\includegraphics[width=0.43\linewidth, trim=0cm 0cm 0cm 0cm, clip=true]{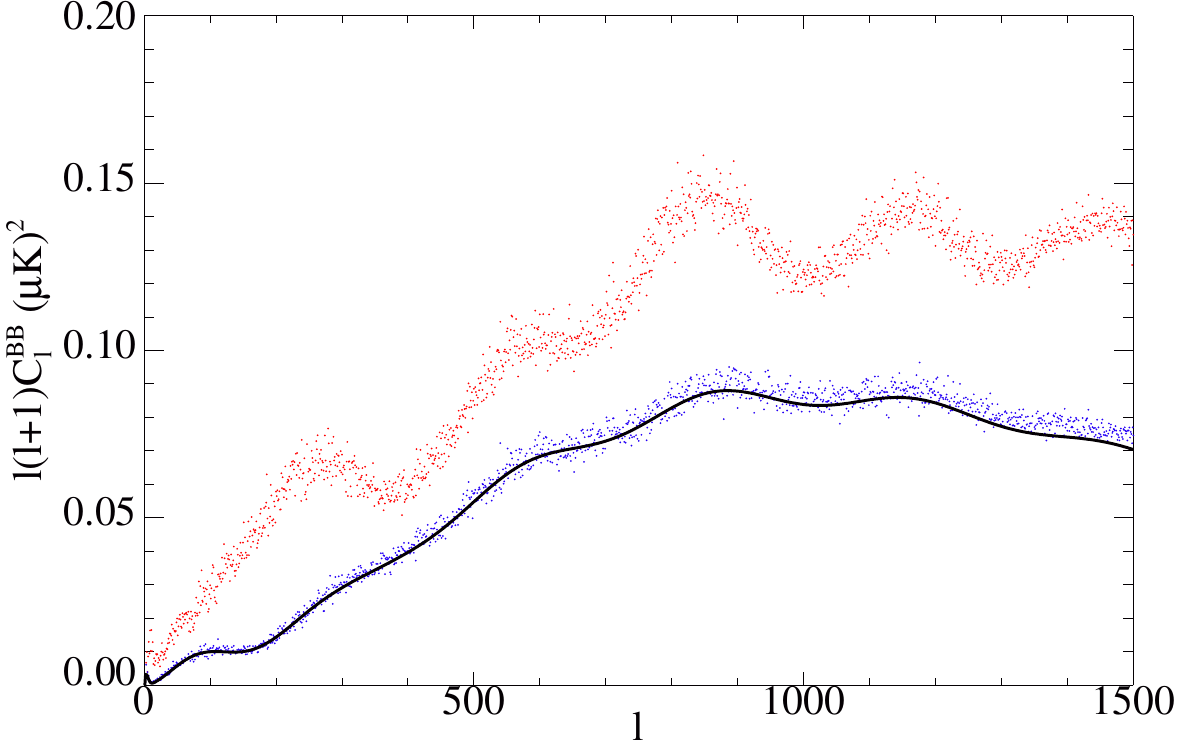} &
\includegraphics[width=0.4143\linewidth, trim=0cm 0cm 0cm 0cm, clip=true]{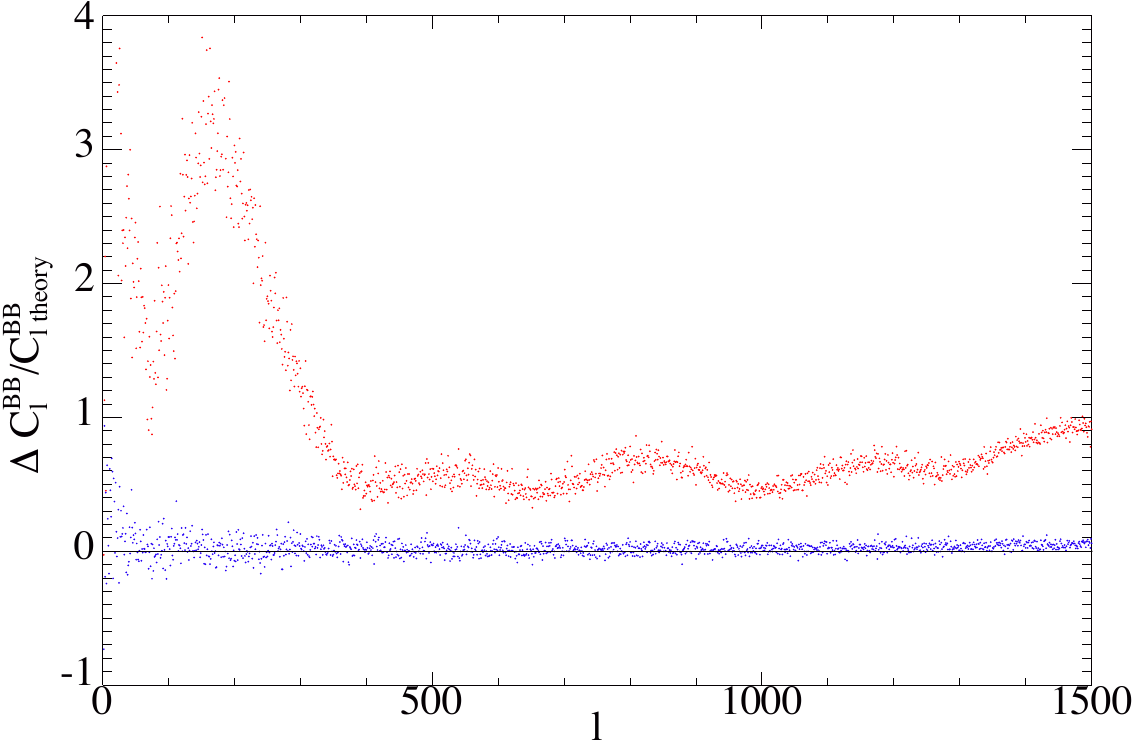} \\
\includegraphics[width=0.43\linewidth, trim=0cm 0cm 0cm 0cm, clip=true]{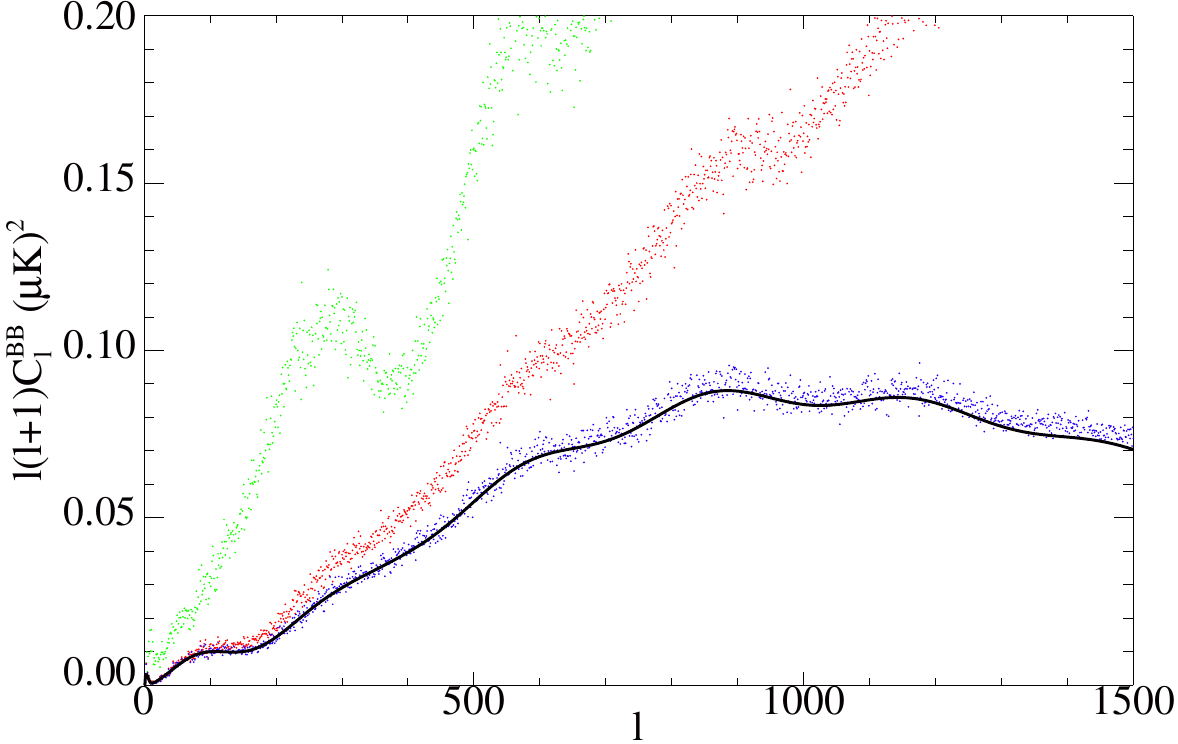} &
\includegraphics[width=0.4143\linewidth, trim=0cm 0cm 0cm 0cm, clip=true]{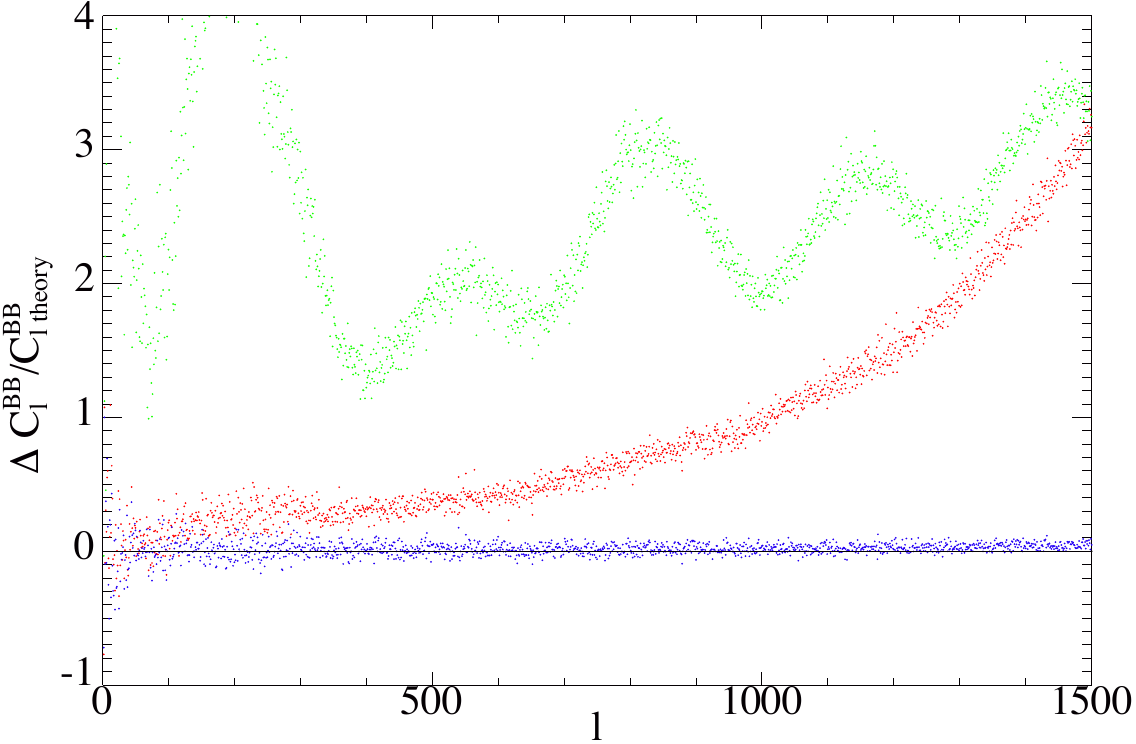} \\
\includegraphics[width=0.43\linewidth, trim=0cm 0cm 0cm 0cm, clip=true]{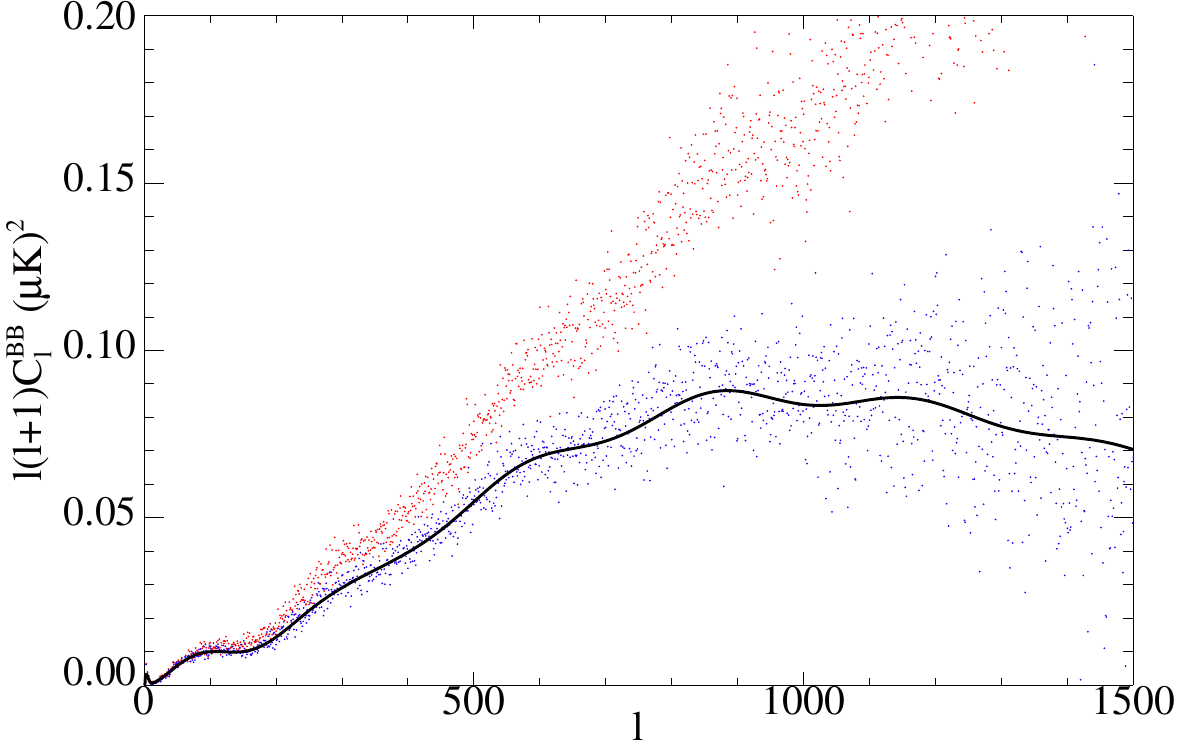}&
\includegraphics[width=0.4143\linewidth, trim=0cm 0cm 0cm 0cm, clip=true]{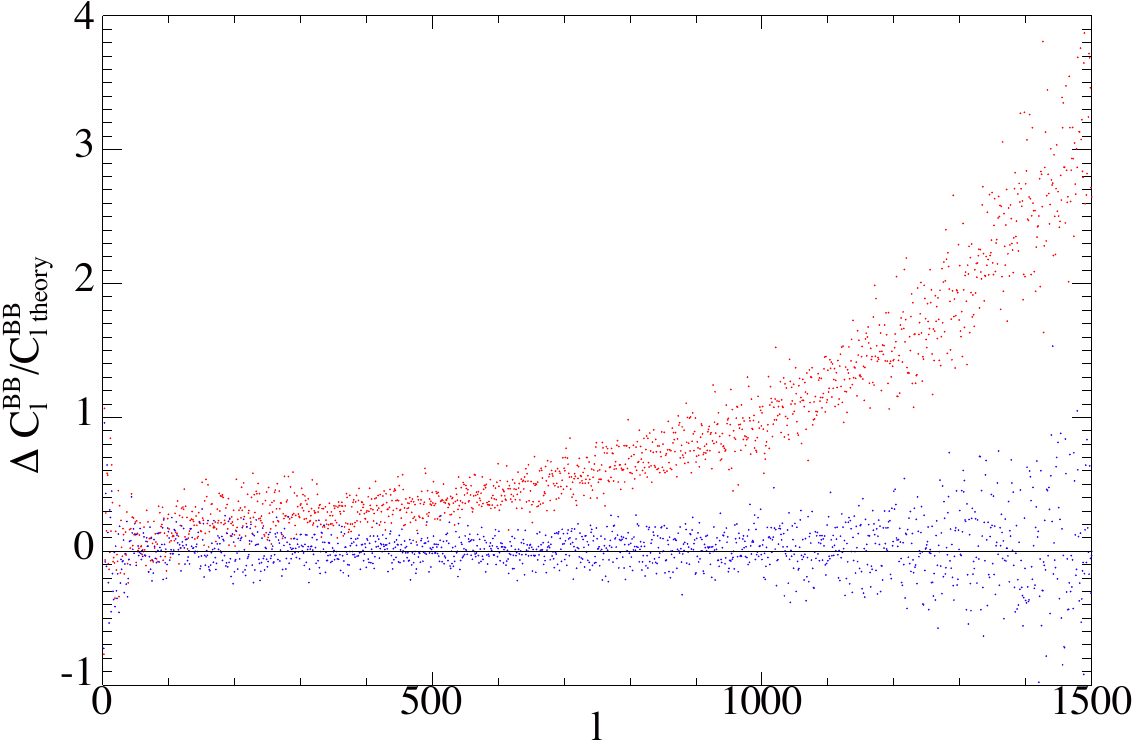} \\
\includegraphics[width=0.43\linewidth, trim=0cm 0cm 0cm 0cm, clip=true]{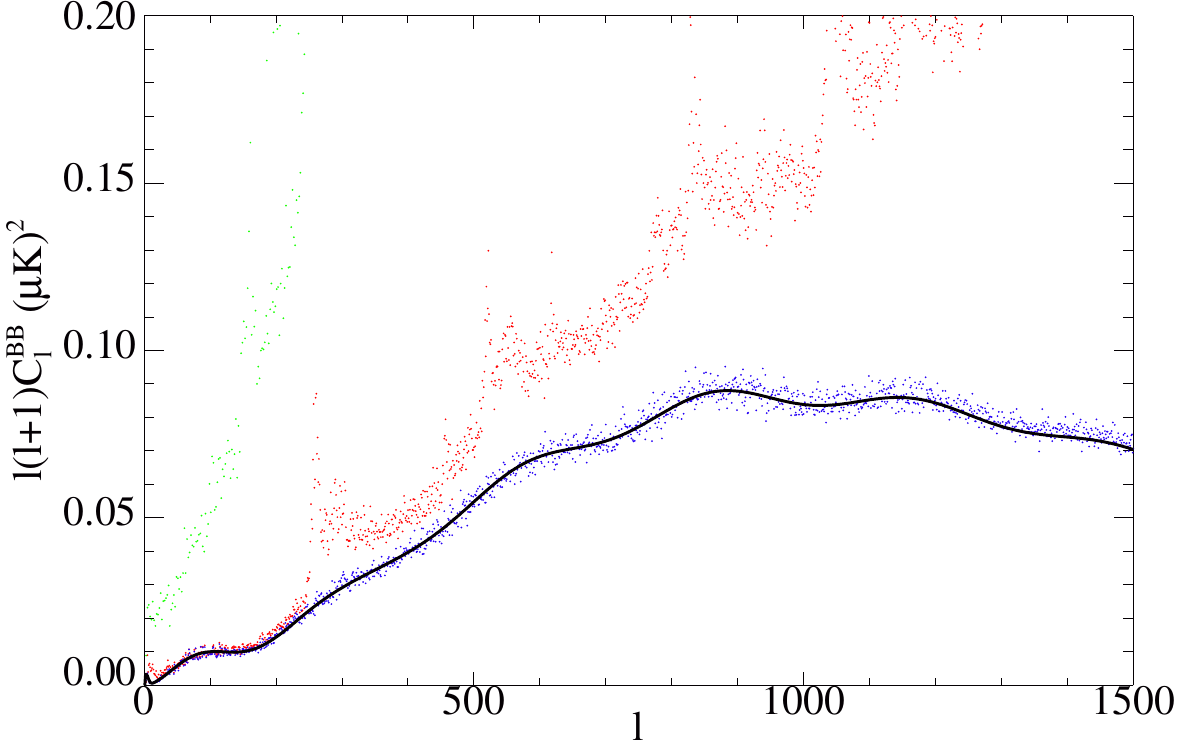} &
\includegraphics[width=0.4143\linewidth, trim=0cm 0cm 0cm 0cm, clip=true]{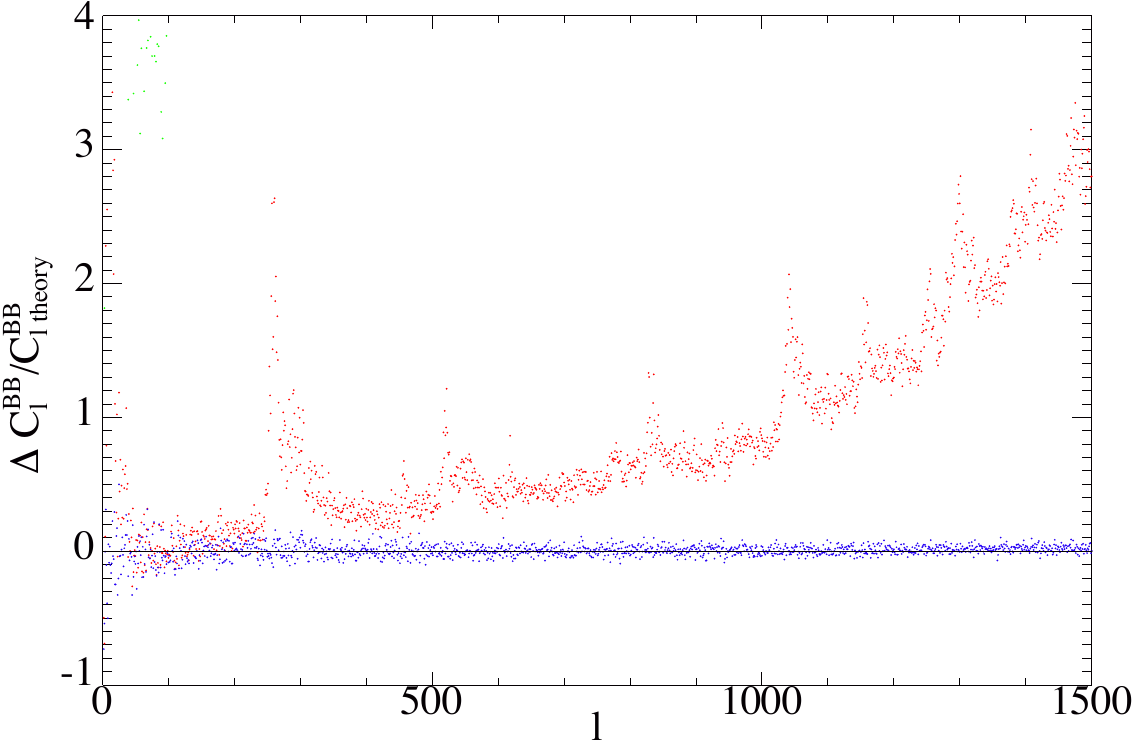} \\
\end{tabular}
\caption{{\color{black}We show the recovered $B$-mode power spectrum for the
  satellite-like experiment as an example of the map-making algorithm
  with a extended range of $\psi_{\rm t}$ angles, see Section
  \ref{sec:psi_t_ext}. The left column shows the power spectra and the
  right the fractional error for the simulations. The red dots show the results
  where a binned map is made and the blue the result when our
  algorithm is used, the input $B$-mode power spectra is shown in black.
{\it First row:} the results from simulation 1, see Section \ref{sec:sim1}, where we
  consider an experiment without a HWP and include differential gain
  and pointing systematics in the TOD. As there is no HWP we cannot apply our method to
  remove differential ellipticity so it is not included here. {\it
    Second row:} the results from simulation 2, see Section
  \ref{sec:sim2}, where we consider an experiment with a HWP and
  included differential gain, pointing and ellipticity systematics in
  the TOD. The green
  dots show the recovered power spectra where a binned map is made and
  the TOD was created without considering a HWP to demonstrate the
  benefit of a HWP at mitigating systematics. {\it Third row:} the
  recovered power spectra for simulation 3, see Section
  \ref{sec:sim3}. This is similar to simulation 2 but with noise
  included in the TOD. {\it Fourth row:} the
  recovered power spectra for simulation 4, see Section
  \ref{sec:sim4}. This is similar to simulation 2 but with the CMB
dipole included in the sky model and a lowed level of differetial gain,
see Section \ref{sec:sim4}. The green
  dots show the recovered power spectra where a binned map is made and
  the TOD was created without considering a HWP to demonstrate the
  benefit of a HWP at mitigating systematics.}
\label{fig:epic_b_mode}}
\end{center}
\end{figure*}

\subsubsection{Simulation 2: HWP included, no noise}\label{sec:sim2}

We simulate a noise-free TOD from a detector pair that suffers from a
differential gain of the two detectors of 1\%, a differential pointing
of 0.25 arcmin which is 3.5\% of the 7 arcmin beam. The differential
ellipticity is created using the beam described by equation
\eqref{eq:gaus_beam}. Figure \ref{fig:epic_b_mode} shows the recovered
$B$-mode power spectrum when a simple binned map is made and a map
using the algorithm described in Section \ref{sec:psi_t_ext}. We
included the terms $\tilde{S}_{0, 0}$, $\tilde{S}_{0, 1}$ and
$\tilde{S}_{0, 2}$ in equation~\eqref{eq:2d_for} to account for the
differential gain, pointing and ellipticity. Figure
\ref{fig:epic_b_mode} shows that the bias created from the systematics
has been removed. We also present the result when the same experiment
is used but with no HWP present and when a binned map is made. This
illustrates how a HWP can partially mitigate systematics. It also
demonstrates that even with this mitigation, further systematic
removal would be required.

\subsubsection{Simulation 3: HWP included, noise included}\label{sec:sim3}

We simulate a noisy version of the TOD used in simulation 2 described
in \ref{sec:sim2}. We include noise in the TOD of 1 ${\rm \mu K
  \sqrt{s}}$. This is optimistic for a CMB experiment. However it is
chosen so that the recovered $B$-mode power is easily detected with
one detector pair. The algorithm can easily deal with noise as the
noise simply propagates through the matrix operation to the map in the
same way as it does in a binned map-making scheme. The algorithm,
however, does increase the noise on the recovered $Q$ and $U$
measurements and it also increases the covariance of the $Q$ and $U$
estimates. This increase depends on the scan strategy: the more
crossing angles, the lower the increase in noise. The limiting cases of
this are (1) an ideal scan strategy, where the noise increase is zero	
and (2) where the matrix in equation \eqref{eq:2d_for} is singular, in
which case the effective increase in the noise is infinite. With noise
simulations we have shown that the noise increase for the EPIC scan and
this HWP set up is 12\%. Fig.~\ref{fig:epic_b_mode} shows the
recovered $B$-mode power spectrum when a simple binned map is made and
a map using the algorithm described in Section \ref{sec:psi_t_ext} and
where we have removed the noise bias in each case. Again we included
the terms $\tilde{S}_{0, 0}$, $\tilde{S}_{0, 1}$ and $\tilde{S}_{0, 2}$ in
equation~\eqref{eq:2d_for} to account for the differential gain, pointing and
ellipticity. Fig.~\ref{fig:epic_b_mode} shows that the bias created
from the systematics has been removed even in the presence of noise.

{\color{black}
\subsubsection{Simulation 4: HWP included, no noise, CMB dipole included}\label{sec:sim4}

The CMB dipole can in principle leak to polarisation through the
systematics considered in this paper and this effect can be very
large.  Here, we test if the map-making algorithm can remove such a
level of leakage. For an experiment where the large scale modes are
not filtered out at the time stream level, the leakage from the CMB
dipole could be problematic. Experiments of this type do have the
benefit of being able to calibrate their detectors using
 the CMB dipole \citep{2015arXiv150201587P}.  With this
benefit in mind one would expect the differential gain for such an
experiment to be lower than for ground-based experiments.  To reflect
this effect we lower the level of differential gain in the simulations
for this section. We simulate a noise-free TOD from a detector pair
that suffers from a differential gain of the two detectors of
0.2\%. The other two systematics were kept the same as in simulation 2
in Section \ref{sec:sim2}.  Figure \ref{fig:epic_b_mode} shows the
recovered $B$-mode power spectrum when a simple binned map is made and
a map using the algorithm described in Section \ref{sec:psi_t_ext}. We
included the terms $\tilde{S}_{0, 0}$, $\tilde{S}_{0, 1}$ and
$\tilde{S}_{0, 2}$ in equation~\eqref{eq:2d_for} to account for the
differential gain, pointing and ellipticity. Figure
\ref{fig:epic_b_mode} shows that the bias created from the systematics
has been removed. We also present the result when the same experiment
is used but with no HWP present and when a binned map is made. This
illustrates how a HWP can partially mitigate systematics. It also
demonstrates that even with this mitigation, further systematic
removal would be required.  }

\subsection{Limited $\psi_{\rm t}$ coverage algorithm with a balloon-like experiment}\label{sec:balloon}

We test the limited $\psi_{\rm t}$ range form of our map making
algorithm on a balloon like experiment. We use the LSPE scan strategy
\citep{2012SPIE.8446E..7AA} --- see Fig.~\ref{fig:hit_maps} for the hit
map of the LSPE scan strategy. This is a typical balloon-like scan
strategy where the gondola rotates rapidly to cover the sky. This
provides good sky coverage (25\% for LSPE). However, the pixels are
always scanned in a similar direction. Although this can make
systematic mitigation problematic, we show here that this problem can
be avoided with a suitable map-making algorithm. Instead of using the
range of crossing angles to accurately characterise the systematic and
therefore remove it, here we assume that the systematics are slow
enough functions of $\psi_{\rm t}$ that in the small range of
$\psi_{\rm t}$ probed by the instrument, the combined effect can be
described by a few Legendre polynomials. The maximum range of
$\psi_{\rm t}$ for all the pixels in the LSPE scan strategy is
$\approx$0.5 rads. Figure \ref{fig:leg_demo} shows that with this small
range the spin-2 systematic (the fastest changing systematic we
consider) can be recovered to a fraction of a percent with just the
first 3 Legendre polynomials.

In the simulation we use the same beam shape as used in
\citet{2014MNRAS.442.1963W} --- see Figure 1 of that paper, which has a
FWHM of 1.5\degree. This beam is a simulation of the beam planned to
be on board LPSE. We simulate TODs from a detector pair ---  one dectector
has the same beam as the other but rotated by $\pi/2$ to provide a
differential ellipticity. LSPE will achieve the required angle
coverage by using a stepped HWP. We simulate this by rotating the
polarisation sensitivity of the beam. We step the HWP by $\pi/8$ every
hour in the simulation. We simulate the LSPE scan strategy for 15
days, each day the telescope performs scans of constant elevation. We
then change the elevation daily.

Unlike in the satellite-like experiment, simulating the asymmetric beam
for polarisation is feasible. Even though the asymmetry of the
polarised beam will in principle create a bias that our algorithm does
not remove, the resulting bias is considerably smaller than the
temperature leakage. The simulation performed here is a good
demonstration of this as we simulate the systematic errors such as
differential pointing completely. However, we only remove the
resulting temperature to polarisation leakage and we do not account
for the resulting polarisation to polarisation errors.

As the amount of simulated data is much smaller for this low
resolution balloon-like experiment, we do not have to make the same
approximations as we do in the satellite-like experiment. We perform a
pixel based integration of the beam multiplied by the CMB sky for each
TOD element in the experiment. We simulate differential gain by
multiplying one detector of the pair by 1.01 to create a 1\%
error. Differential pointing is created by changing the pointing
position of the second detector’s beam by 0.05\degree~which is 3\% of
the 1.5\degree~FWHM beam. As described above the differential ellipticity
is created using the simulated beam shown in figure 1 of
\citet{2014MNRAS.442.1963W}. We include noise in the TOD corresponding
to a noise level in the map of 0.1$~\mu$K per 1.5$\degree$~beam. This level of
noise is optimistic for LSPE (which should achieve $\sim7~\mu$K per
1.5$\degree$~beam \citep{2012SPIE.8446E..7CB}). However, as an example
of the algorithms ability to deal with noise, this level of noise is
more than sufficient.

\begin{figure}
\begin{center}
\includegraphics[width=\linewidth, trim=0cm 0cm 0cm 0cm, clip=true]{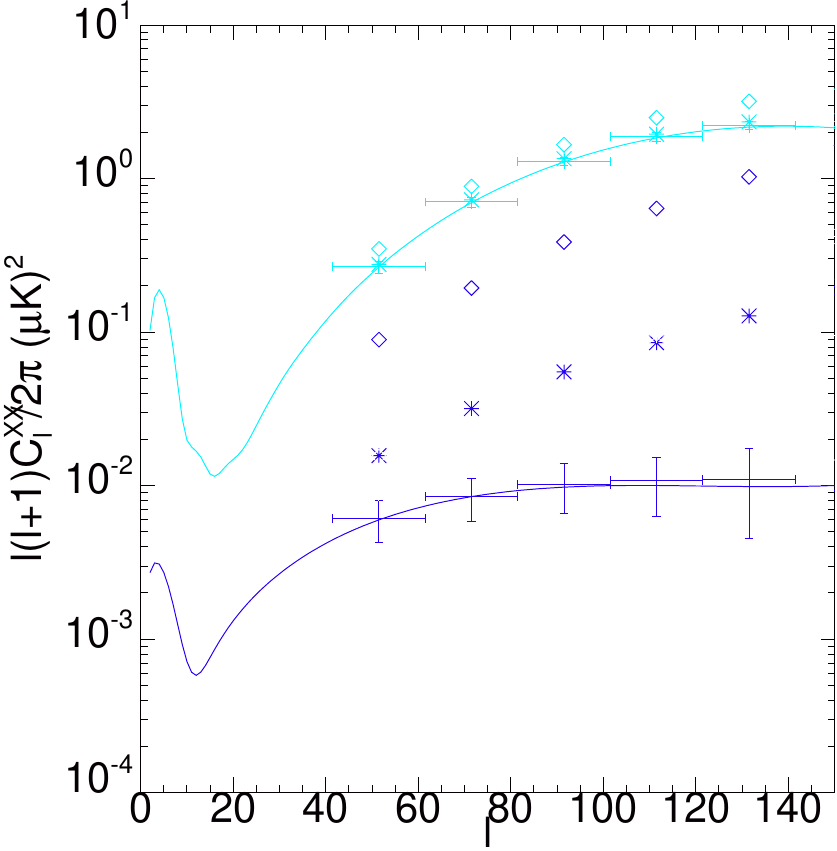} 
\caption{We plot the reconstruction of the $E$-mode and $B$-mode power
  spectrum for the balloon-like experiment, see Section
  \ref{sec:balloon}, as an example of the map-making algorithm for a
  scan strategy that has a limited range of $\psi_{\rm t}$. We show
  the recovered power spectrum averaged over 100 realisations for maps
  created in three ways. The diamonds show the recovered power spectra for a binned map, the stars correspond to including the term $\tilde{S}_{0,0}$ in equation \eqref{eq:2d_forleg}. Finally we show the power spectra including both $\tilde{S}_{0,0}$ and $\tilde{S}_{0,1}$ in equation \eqref{eq:limit_model}. The resulting recovered power spectra and error bars are shown. The bias in the recovered $B$-mode has been reduced by 2 orders or magnitude to less than 5\% of the error bars, which as described in Section \ref{sec:balloon} is for an optimistic LSPE noise.
\label{fig:lspe_sim}}
\end{center}
\end{figure}

Figure \ref{fig:lspe_sim} shows the recovered $E$- and $B$-mode power
spectrum for the balloon-like experiment. As the LSPE scan strategy
only covers 25\% of the sky we use a simple {\color{black}pseudo-$C_\ell$} estimator
\citep{2005MNRAS.360.1262B} to recover the polarised power spectra. We
show the recovered power spectrum averaged over 100 realisations for
maps created in three ways. Unlike in the extensive $\psi_{\rm t}$
range case there is little physical interpretation of the terms
removed in the limited $\psi_{\rm t}$ range case. We are creating
different approximations to the combined effect of the systematic
$f(x)$ from equation \eqref{eq:limit_model}. The first map-making
algorithm we use assumes $f(x){=}0$. This is equivalent to a simply
binned map and the recovered power spectra are shown as the diamonds
in figure \ref{fig:lspe_sim}. The bias from the binned map is obvious.
We can improve this by including the term $\tilde{S}_{0,0}$ in
equation~\eqref{eq:2d_forleg}. This makes the approximation
$f(x){=}{\rm const}$ and reduces the bias by an order of magnitude,
as shown in Fig.~\ref{fig:lspe_sim} as the stars. Finally we improve
this further by assuming that $f(x)$ is a linear function. This is
done by including both $\tilde{S}_{0,0}$ and $\tilde{S}_{0,1}$ in
equation~\eqref{eq:limit_model}. The resulting recovered power spectra
and error bars are shown in Fig.~\ref{fig:lspe_sim}. The bias in the
recovered $B$-mode has been reduced by 2 orders or magnitude to less
than 5\% of the error bars, which as described above are error bars
for an optimistic LSPE noise level.

Figure \ref{fig:lspe_sim} clearly demonstrates that our technique can
remove the bias as a result of temperature to polarisation
leakage. However, this comes at the cost of a noise penalty in the
map. Including the $\tilde{S}_{0,0}$ only and adopting $f(x){=}{\rm
  const}$ incurs a 5\% increase in the noise power with respect to the
binned map. The analysis where $f(x)$ is modelled as a linear function
and both $\tilde{S}_{0,0}$ and $\tilde{S}_{0,1}$ terms are included in
equation \eqref{eq:limit_model} creates an increase in the noise power
of 12\% with respect to a binned map.

\section{Identifying Systematics}\label{sec:identify_sys}

The extensive $\psi_{\rm t}$ range map-making algorithm relies on
being able to remove systematics which have different Fourier modes
in $(\psi_{\rm t},~\psi_{\rm r})$ space to that of the polarisation
signal. As discussed in section \ref{sec:mapmaking_hwp}, the
map-making algorithm requires us to know that the systematics are present
in the TOD so that we can choose to include the correct Fourier modes in
equation \eqref{eq:2d_for} and create maps that are clean of those
systematics. We stress that we do not need to know the exact nature of the
systematic. For example, if the experiment is suffering from 
differential pointing of the detector pair, at no point do we need to
know by how much or in what orientation the beams are
misaligned. Moreover, we do not require a temperature map to remove the
signal. This is an improvement over the method used in
\citet{2014arXiv1403.4302B}. We only need to know that the experiment
is suffering from differential pointing and therefore we know to
include the $\tilde{S}_{0,1}$ term in the analysis. One down side of
using this method to clean systematics is the increase in statistical
noise. Every term included in the analysis increases the statistical
noise of the recovered polarisation maps, and also the cross
correlation of Stokes $Q$ and $U$. The level of the increase is
dependent upon the scan strategy --- the more extensive $(\psi_{\rm t},
\psi_{\rm r})$ coverage the experiment has, the smaller the increase
of noise. We demonstrated this with simulations of the EPIC scan
strategy (see Section \ref{sec:sim_test}). The increase in the noise power
spectrum, going from a binned map to our map-making algorithm accounting for
all three systematics, was 12\%. This would be lower if not all
systematics were considered. With this increase of noise in mind it
would be undesirable to include terms needlessly, but we also do not
want to create a bias by neglecting a term if the systematic is present.

We now turn our attention to a practical process to determine whether
a potential systematic should be removed. We start by making a map of
the systematic. This is done in the same way that the $Q$ and $U$ maps
are made. The inversion of equation \eqref{eq:2d_for} will give us an
estimate for the $\tilde{S}_{n,m}$ terms we included. By taking the
real and imaginary parts of this we can make a map of our
systematics. Figure \ref{fig:sys_maps} shows maps of the systematics
recovered from simulation 3 described in Section \ref{sec:sim3}. We
also show the predictions for each systematic based on prior knowledge,
and finally, the difference between the prediction and recovered systematic
maps. This figure demonstrates how we can accurately recover the
systematics without any prior knowledge of the instrumental
imperfections or of the underlying temperature field.

\begin{figure*}
\begin{center}
\begin{tabular}{c c c c}
&Recovered in map-making&Predicted using simulation inputs&Difference\\
(a)&
\includegraphics[width=\length\linewidth, trim=0cm 0.5cm 0cm 2cm, clip=true]{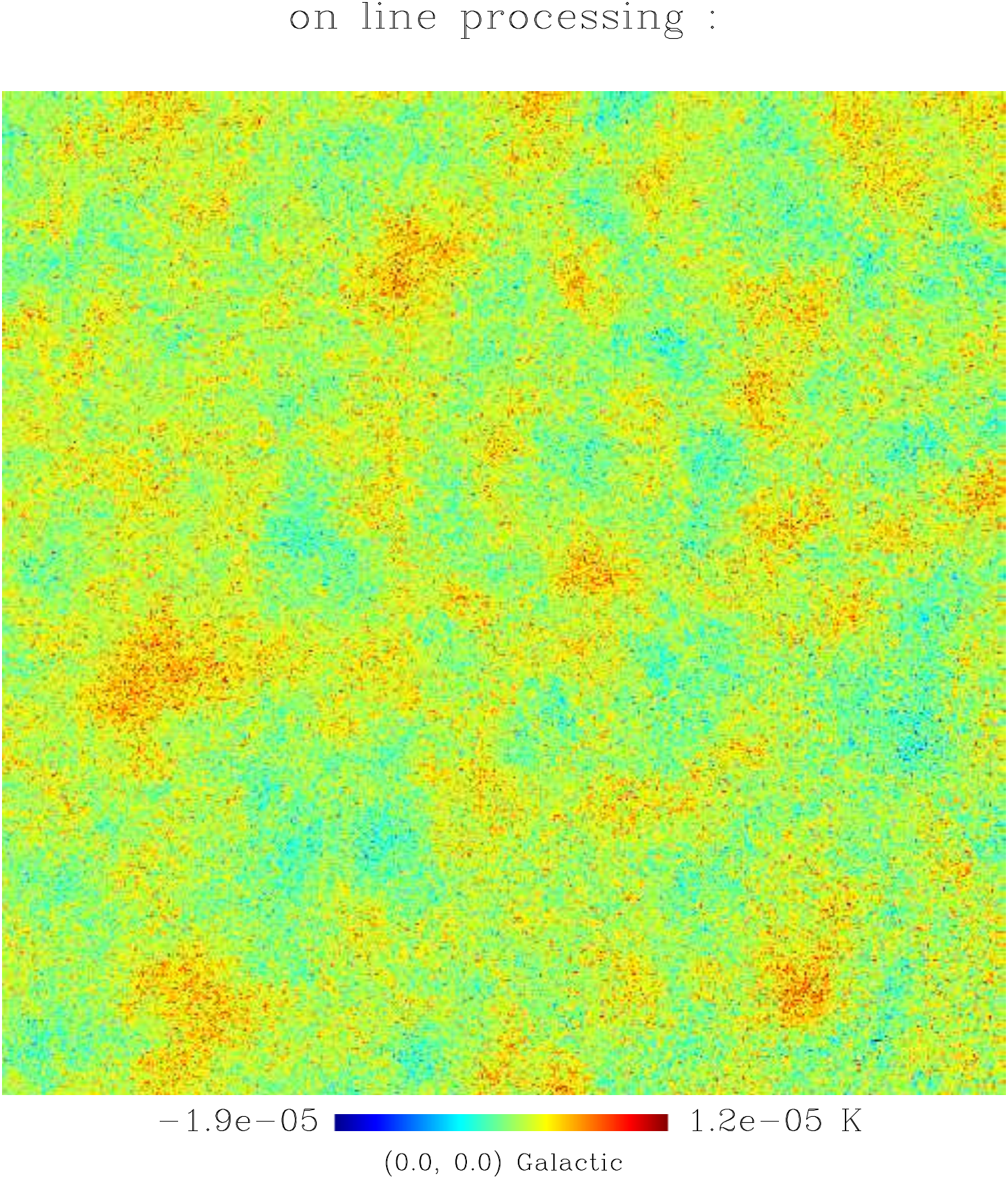} &
\includegraphics[width=\length\linewidth, trim=0cm 0.5cm 0cm 2cm, clip=true]{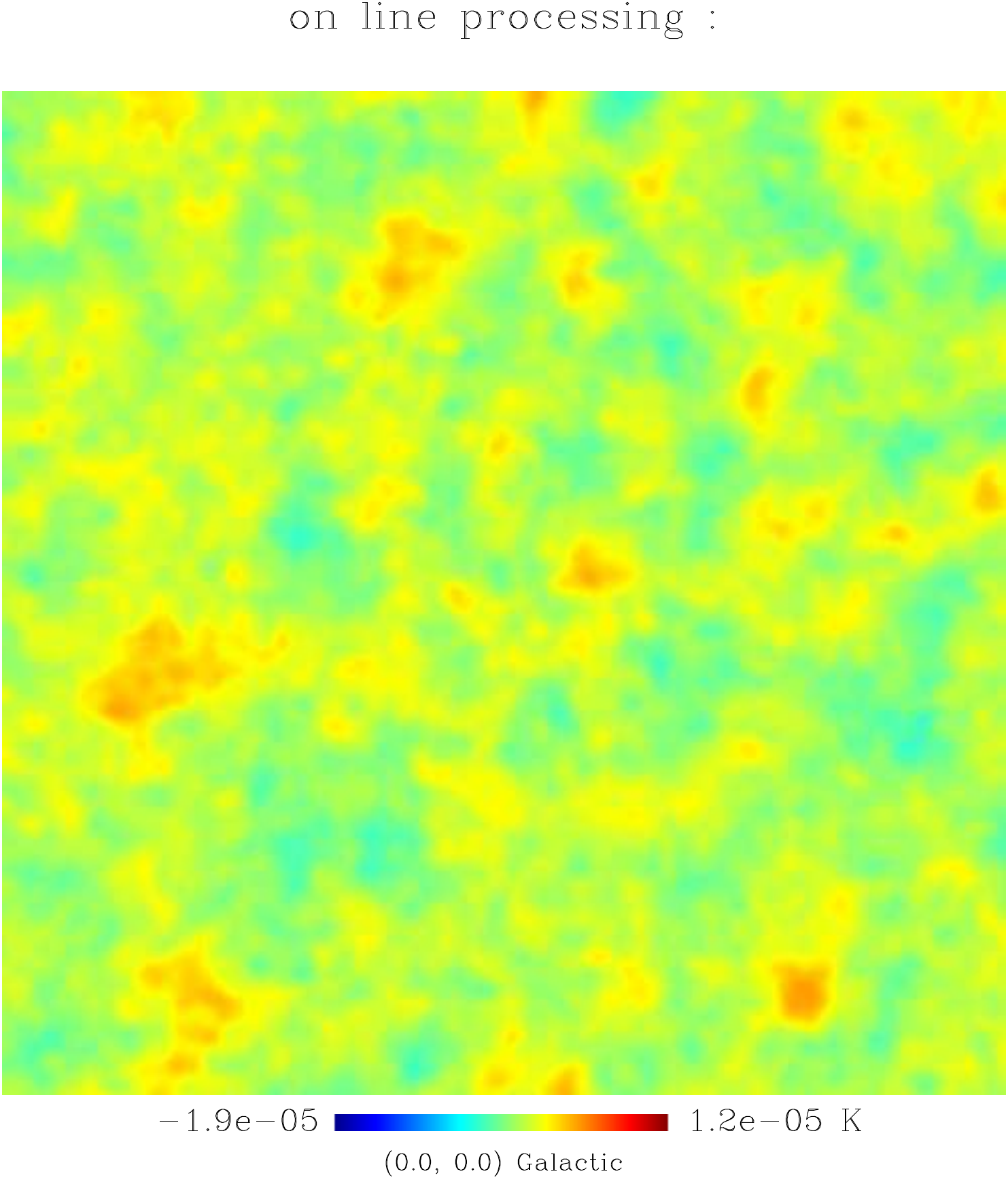}&
\includegraphics[width=\length\linewidth, trim=0cm 0.5cm 0cm 2cm, clip=true]{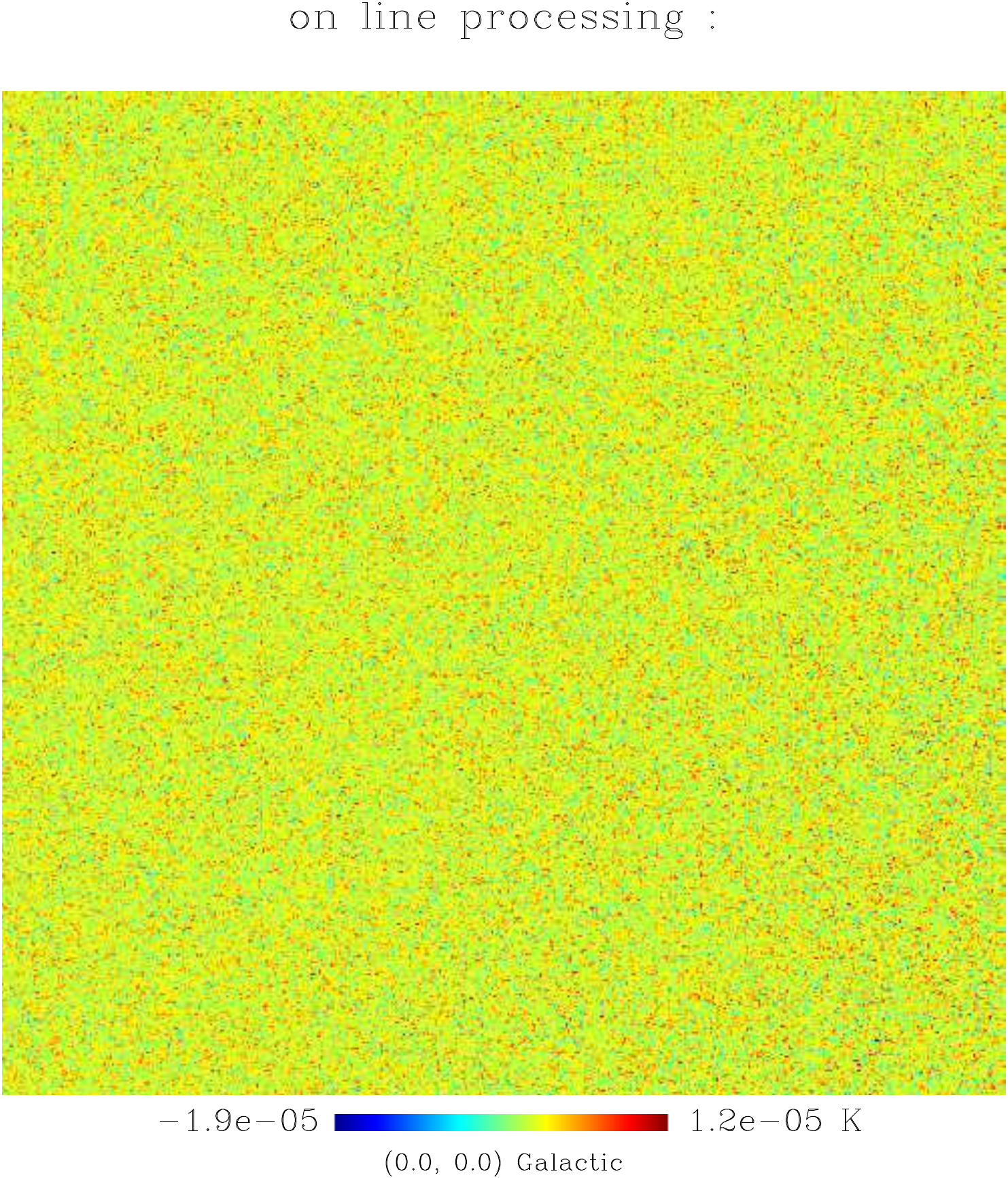}\\
(b)&
\includegraphics[width=\length\linewidth, trim=0cm 0.5cm 0cm 2cm, clip=true]{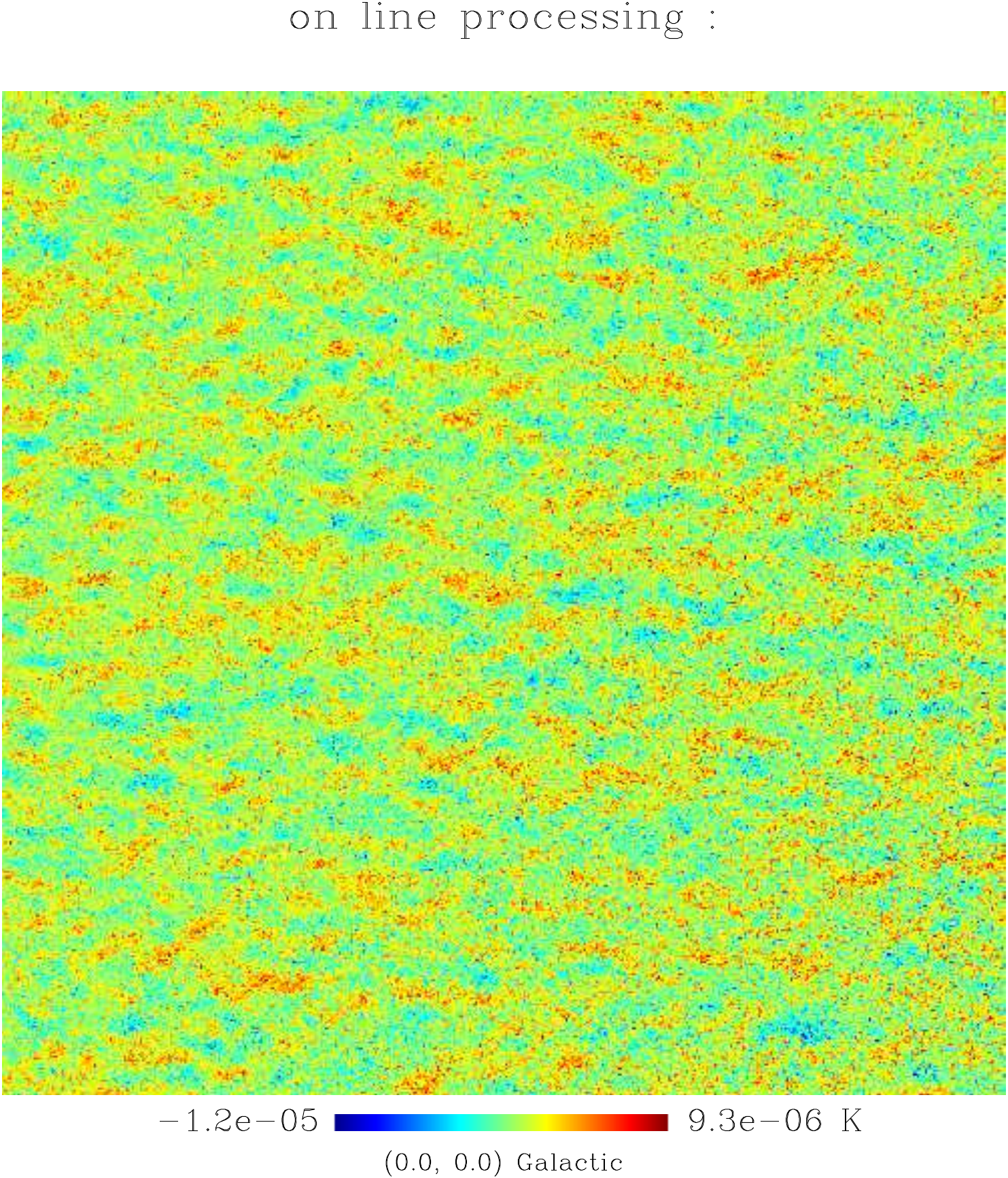}& 
\includegraphics[width=\length\linewidth, trim=0cm 0.5cm 0cm 2cm, clip=true]{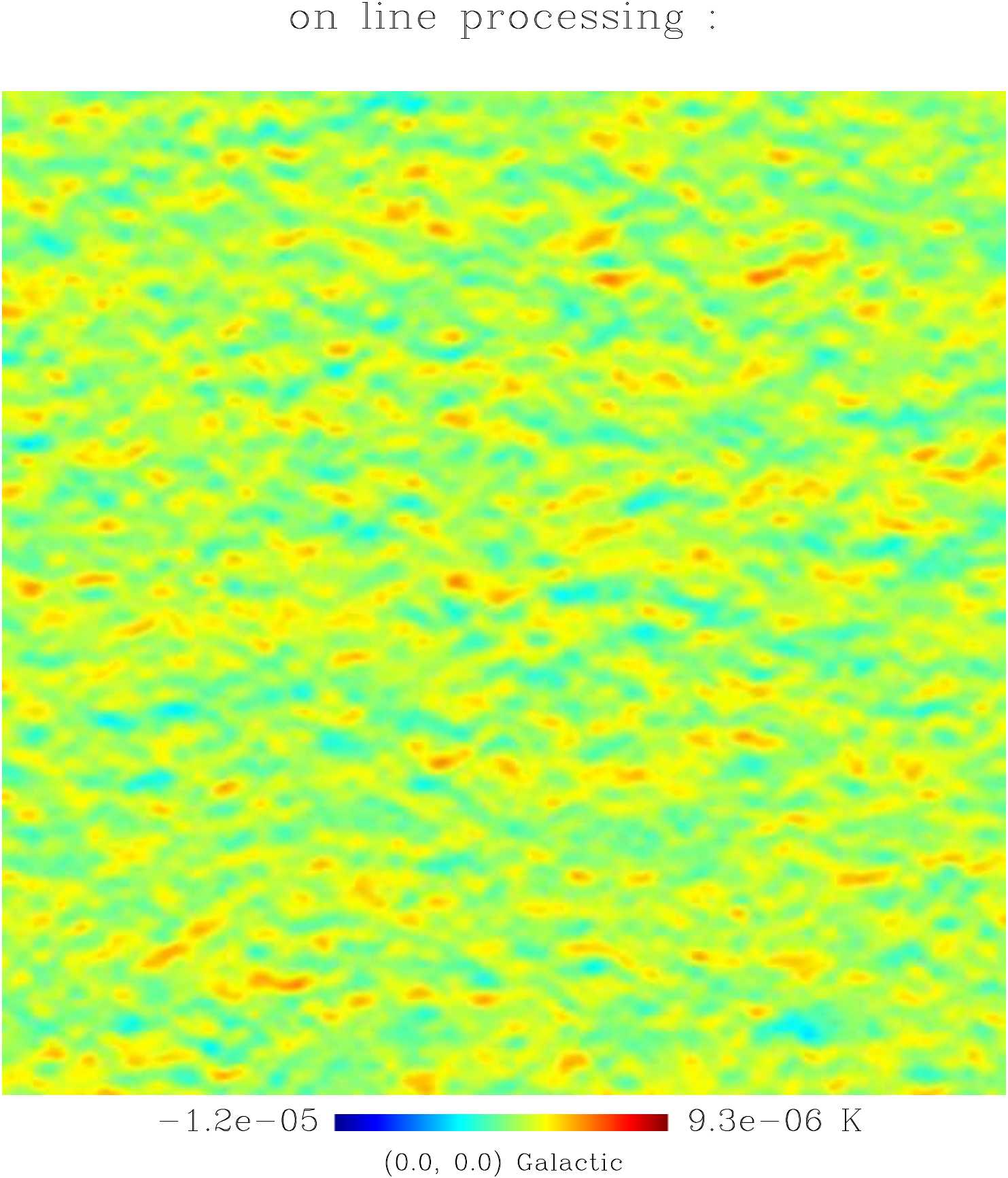}&
\includegraphics[width=\length\linewidth, trim=0cm 0.5cm 0cm 2cm, clip=true]{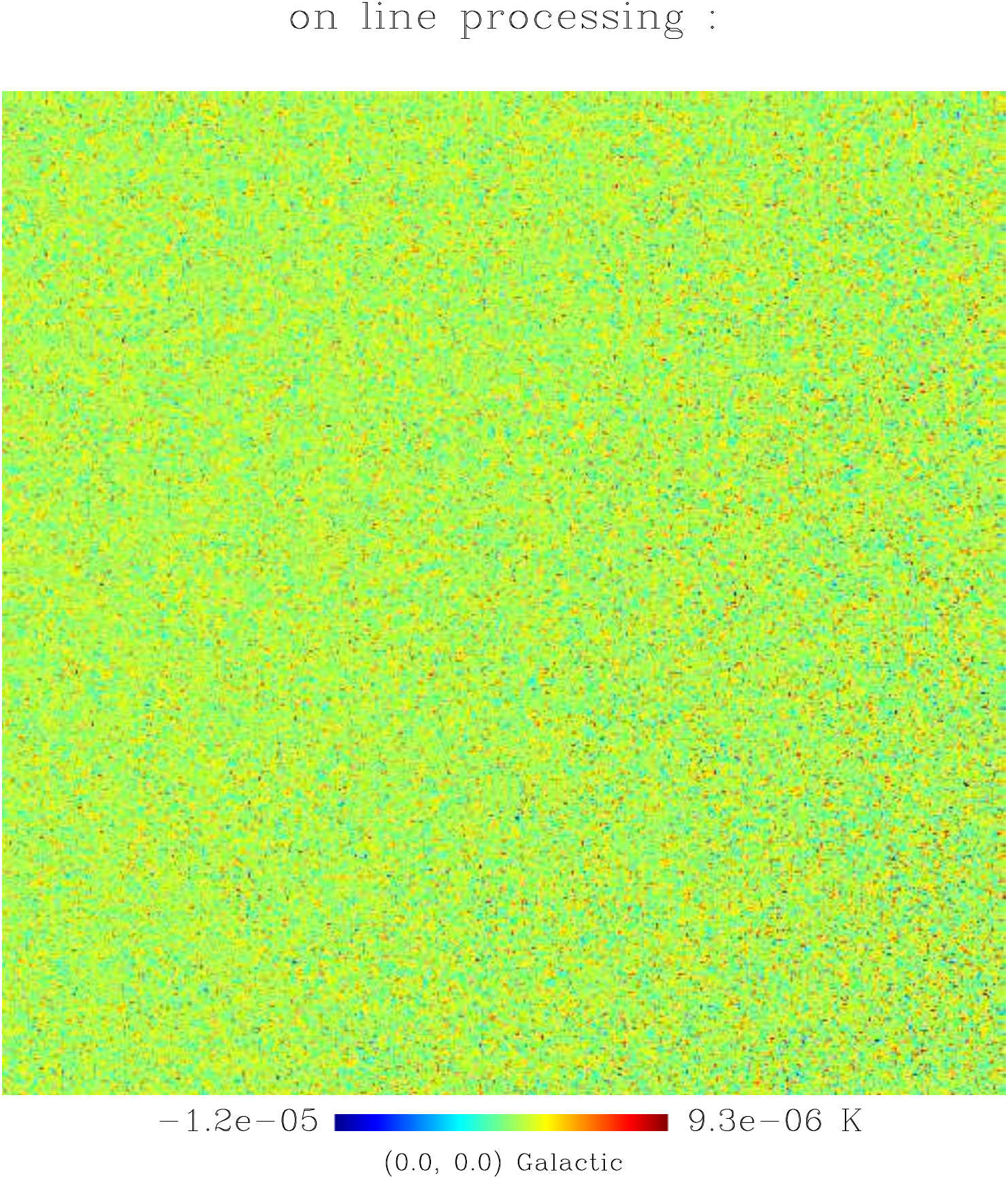}\\
(c)&
\includegraphics[width=\length\linewidth, trim=0cm 0.5cm 0cm 2cm, clip=true]{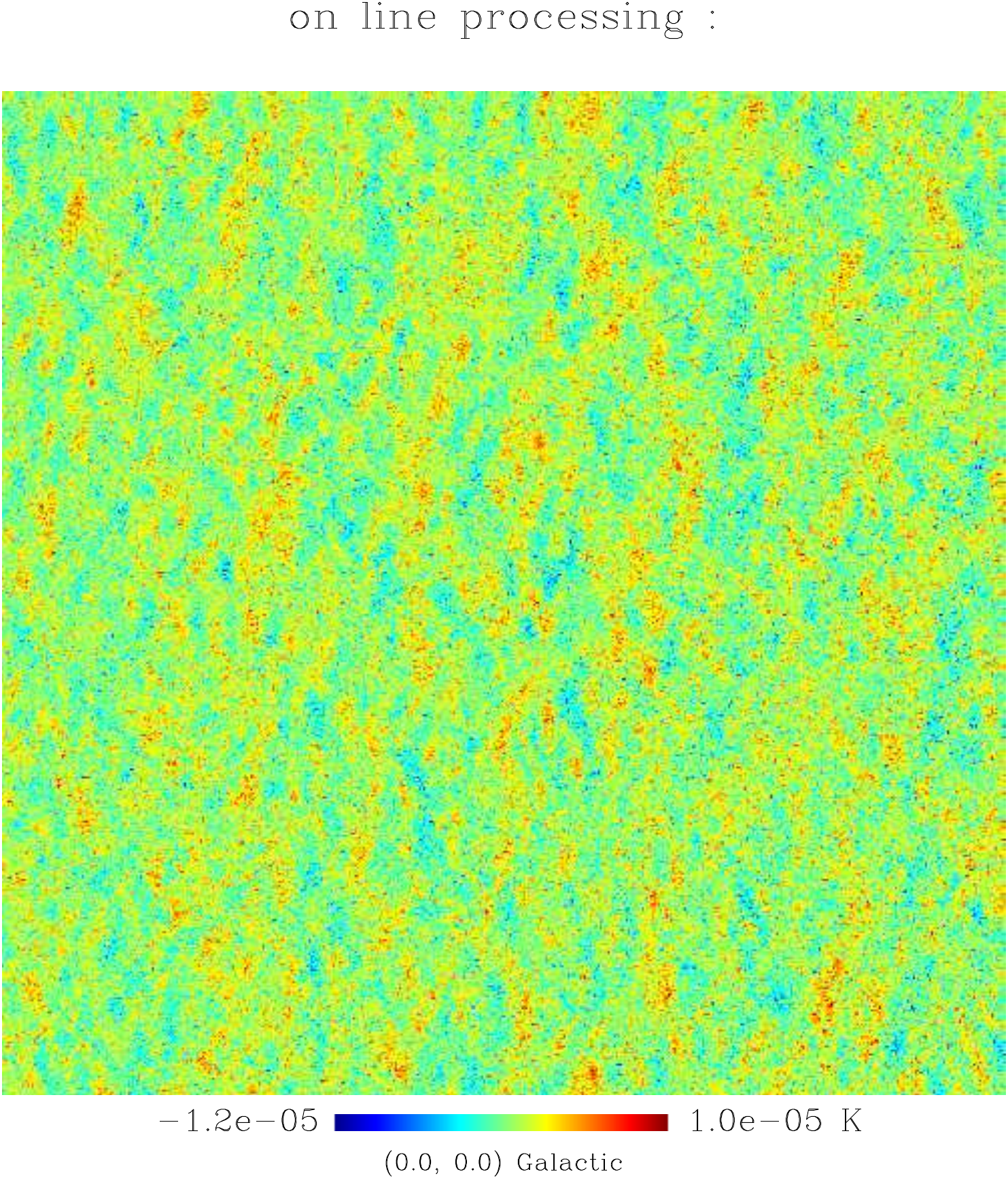} &
\includegraphics[width=\length\linewidth, trim=0cm 0.5cm 0cm 2cm, clip=true]{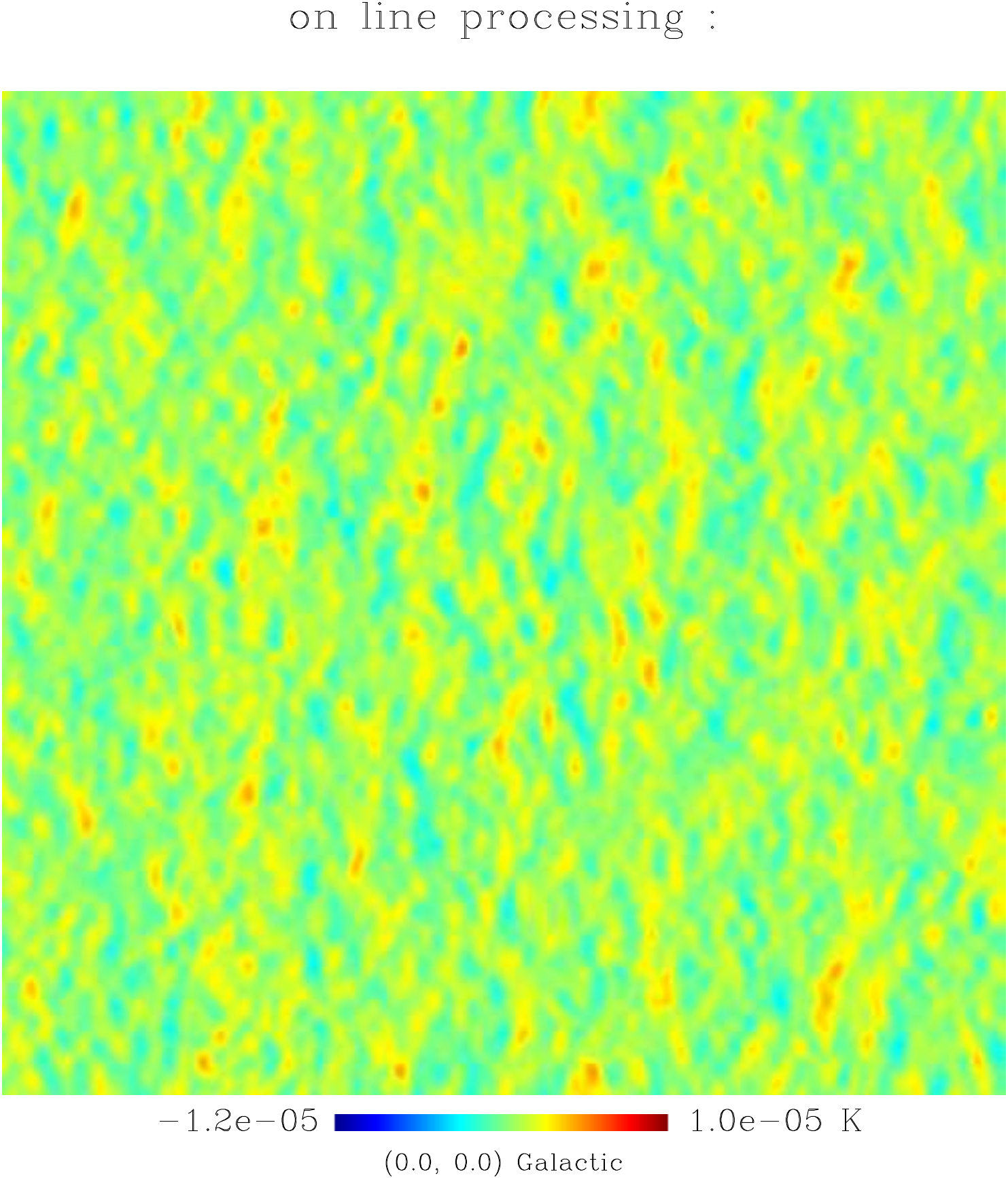}&
\includegraphics[width=\length\linewidth, trim=0cm 0.5cm 0cm 2cm, clip=true]{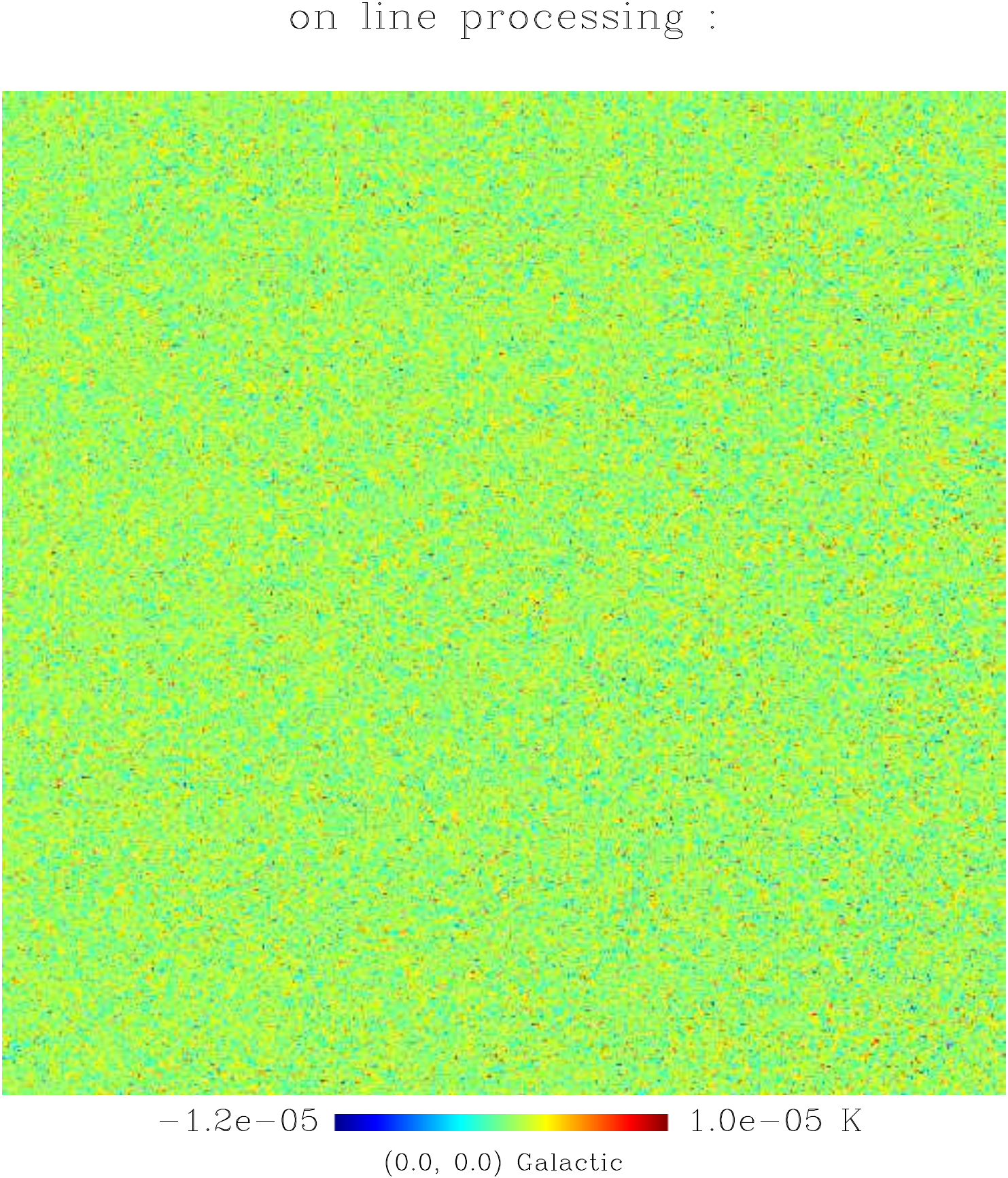}\\
(d)&
\includegraphics[width=\length\linewidth, trim=0cm 0.5cm 0cm 2cm, clip=true]{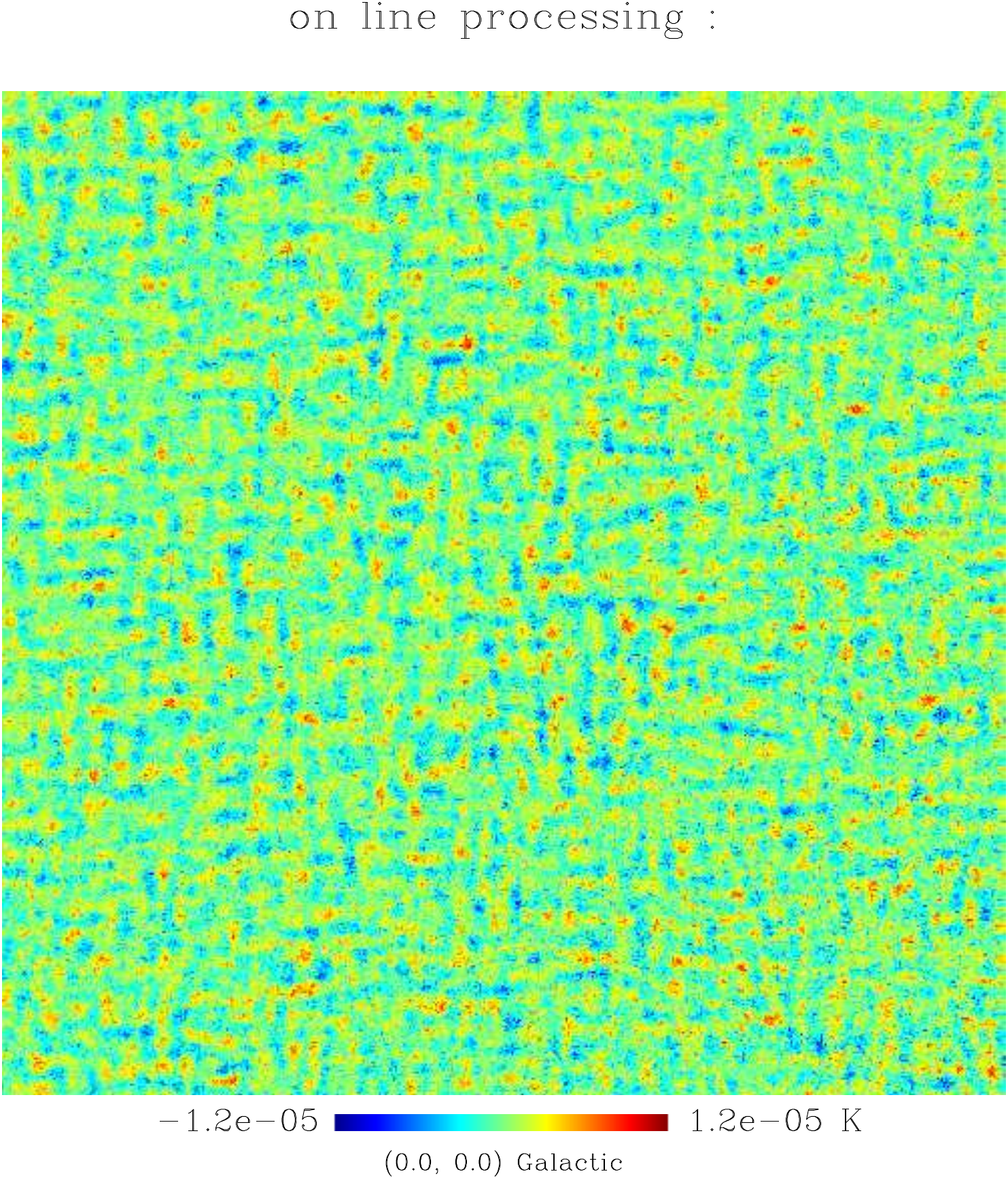} &
\includegraphics[width=\length\linewidth, trim=0cm 0.5cm 0cm 2cm, clip=true]{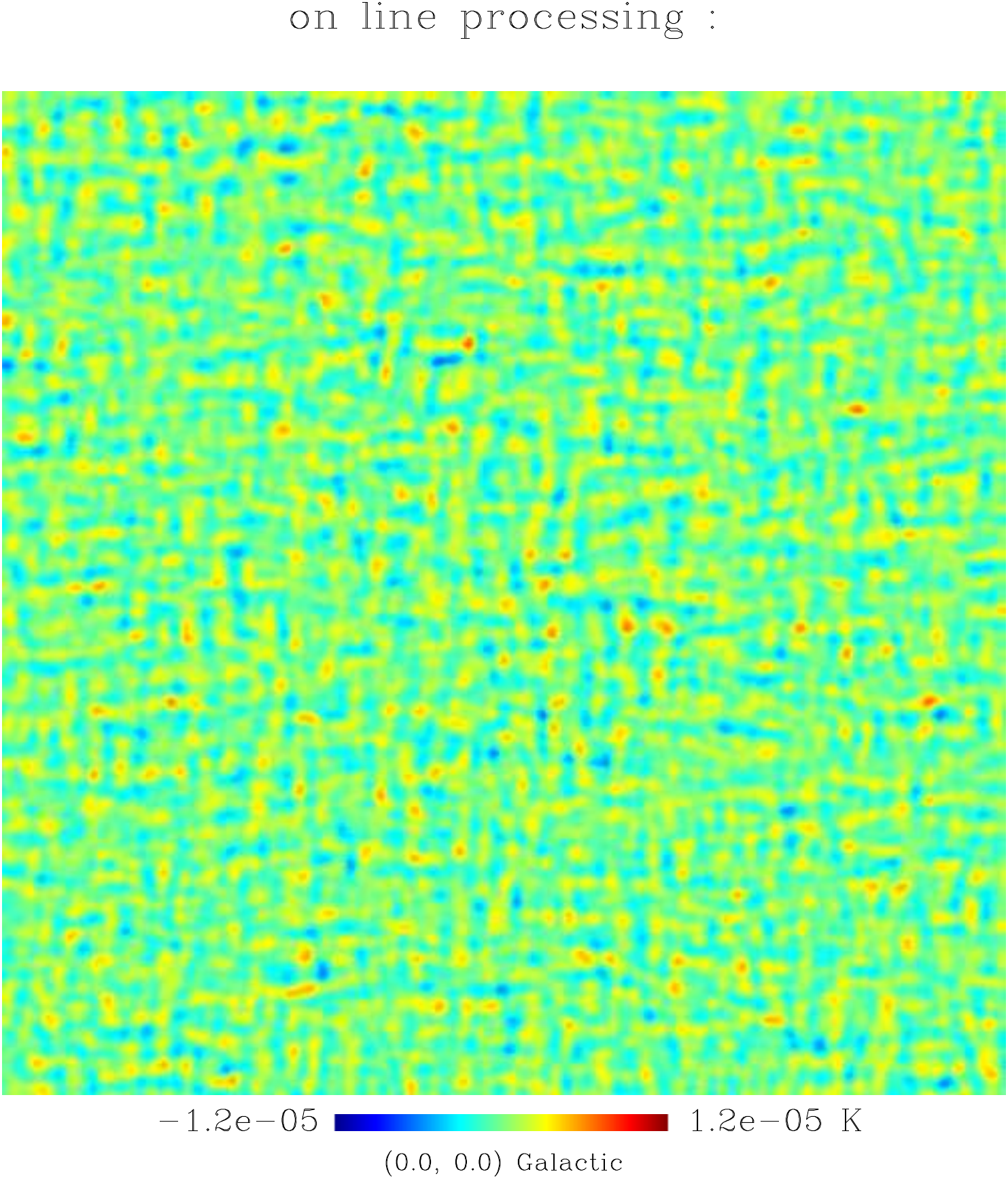}&
\includegraphics[width=\length\linewidth, trim=0cm 0.5cm 0cm 2cm, clip=true]{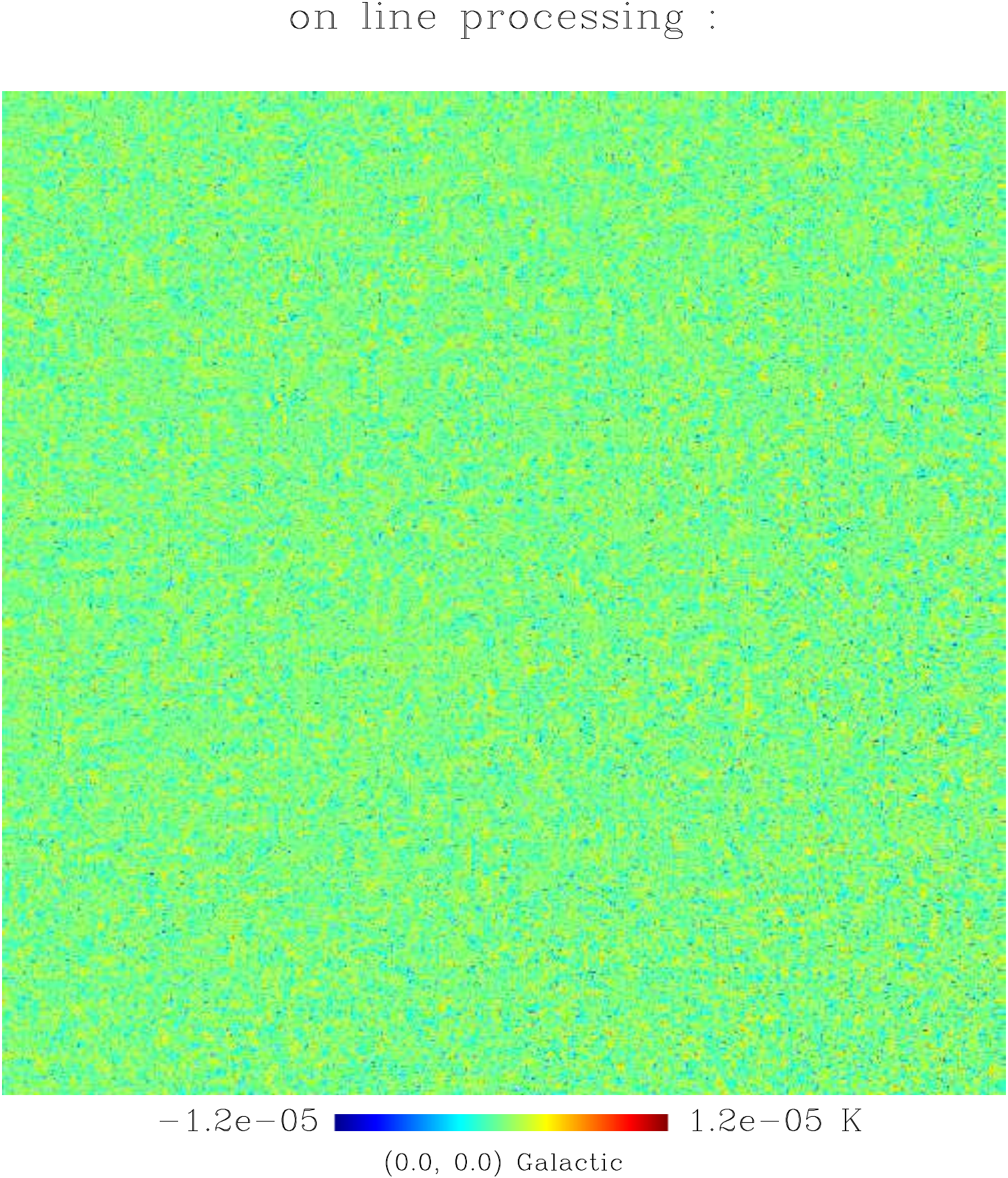}\\
(e)&
\includegraphics[width=\length\linewidth, trim=0cm 0.5cm 0cm 2cm, clip=true]{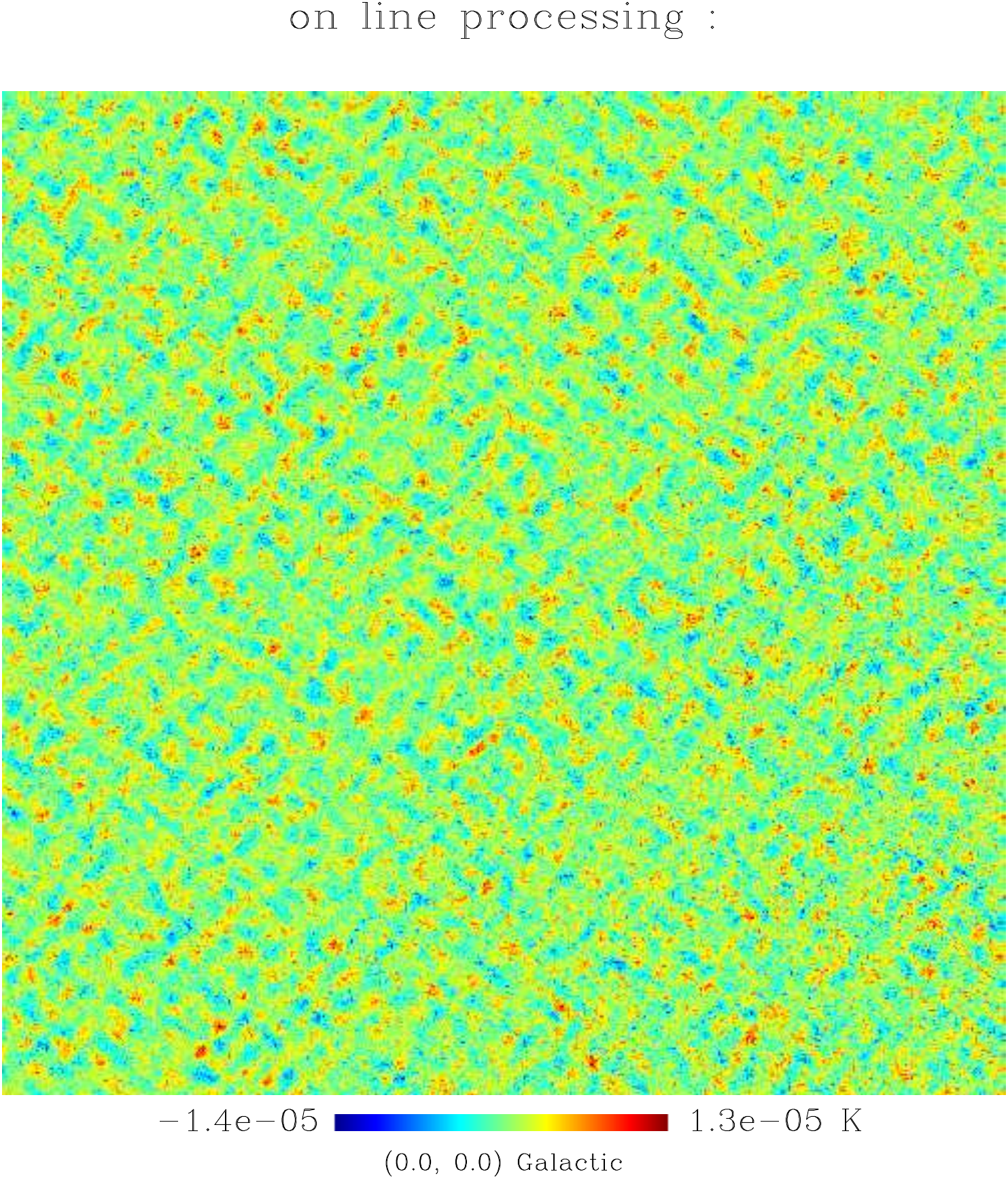} &
\includegraphics[width=\length\linewidth, trim=0cm 0.5cm 0cm 2cm, clip=true]{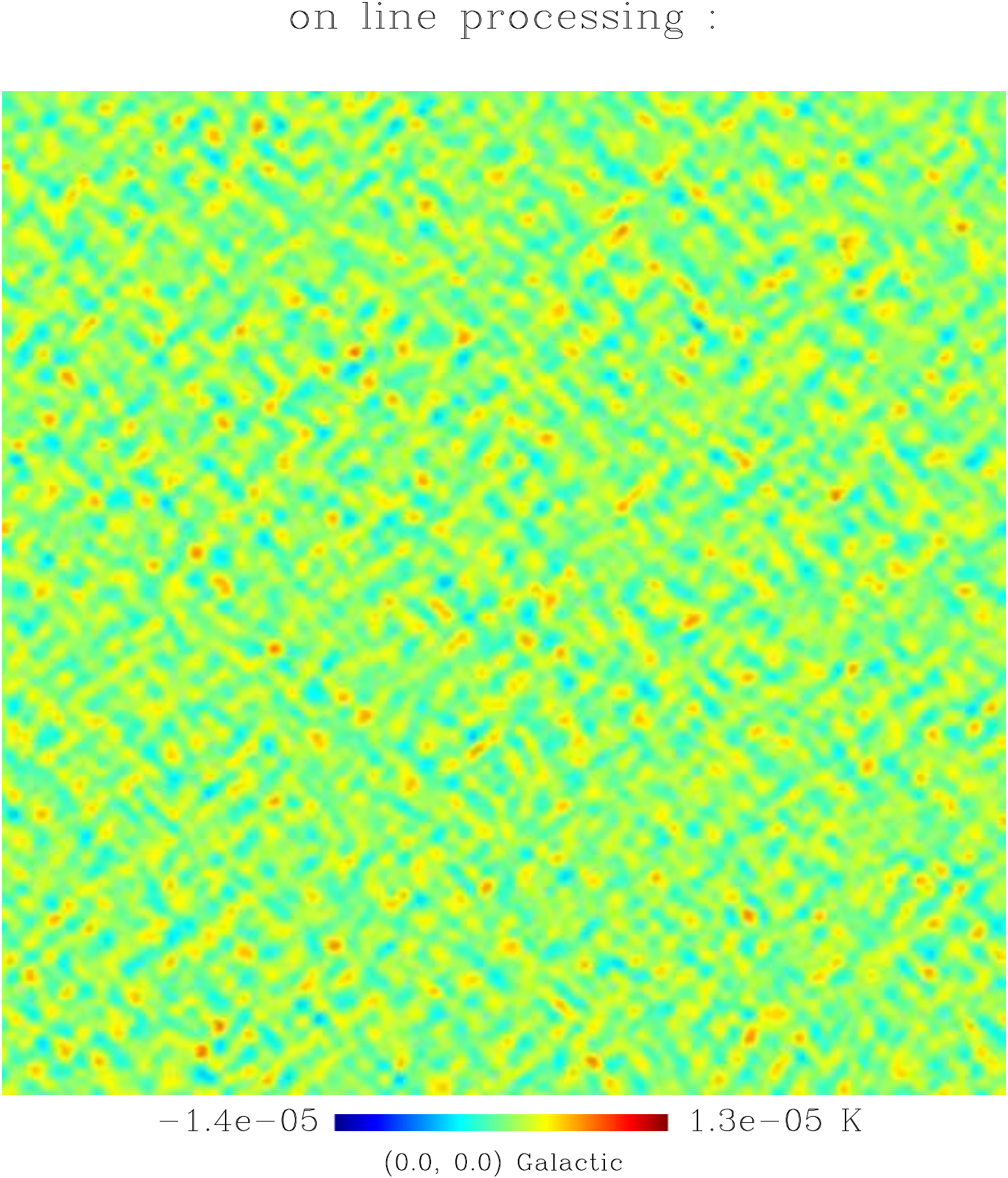}&
\includegraphics[width=\length\linewidth, trim=0cm 0.5cm 0cm 2cm, clip=true]{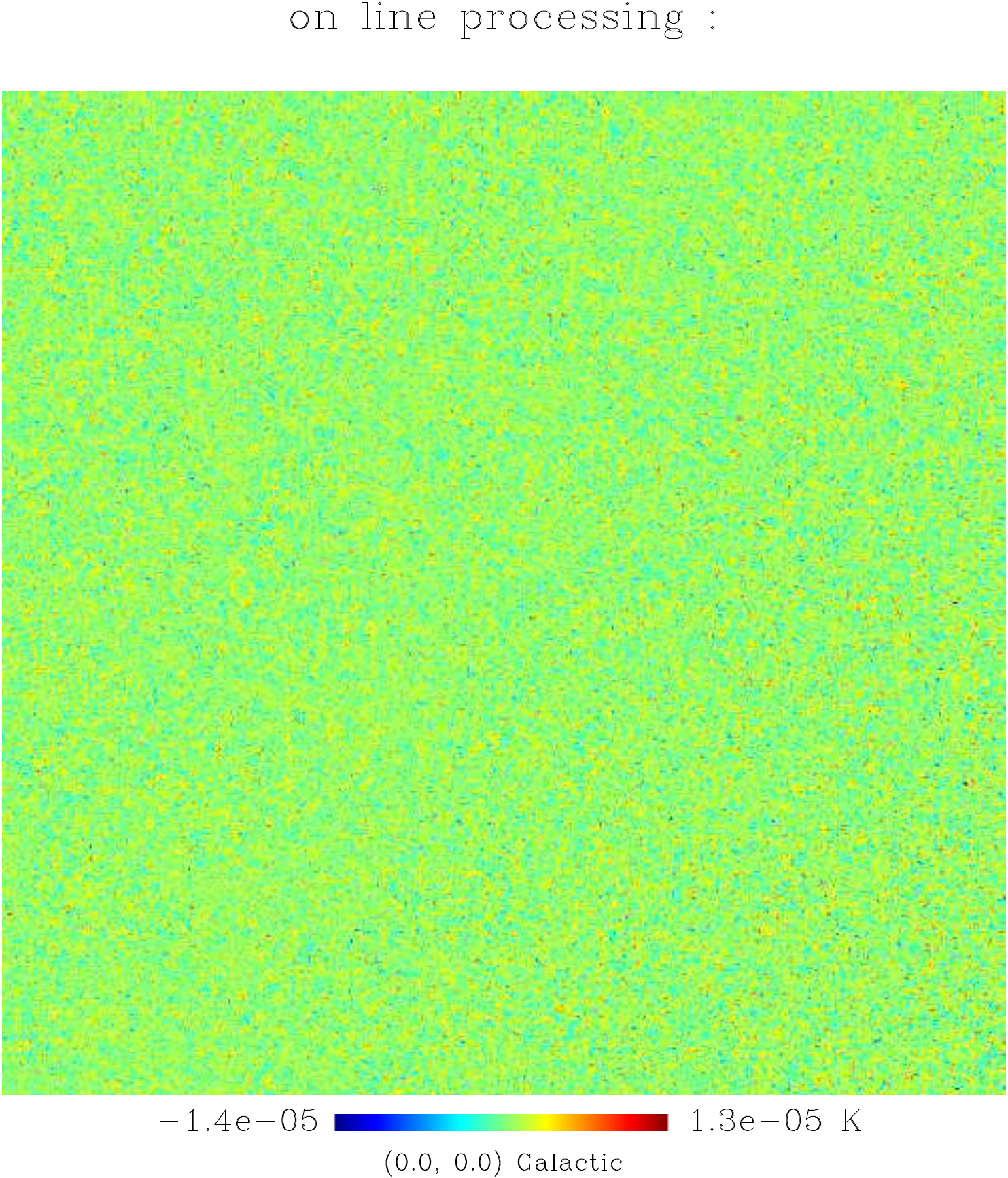}\\
\end{tabular}
\caption{\color{black}Each row shows the systematic from the noisy TOD in simulation 3 described in Section \ref{sec:sim3}. Each section of the sky is a gnomonic projection of a 12.5$\degree$ square patch of sky. We plot the map extracted using the systematic map making method described in Section \ref{sec:mapmaking_hwp} on the left. The middle column displays a prediction for the systematic based on prior knowledge. The right column shows the difference which in each case is compatible with noise. Row (a) shows the results for differential gain ($\tilde{S}_{0,0}$), rows (b) and (c), differential pointing ($\Re[\tilde{S}_{0,1}]$ and $\Im[\tilde{S}_{0,1}]$) and rows (d) and (e) show the differential ellipticity systematic ($\Re[\tilde{S}_{0,2}]$ and $\Im[\tilde{S}_{0,2}]$). 
\label{fig:sys_maps}}
\end{center}
\end{figure*}

The top row of Figure \ref{fig:sys_maps} shows the map of differential
gain. As this is a spin-0 term there is only one non-trivial map to
show. The prediction is created by convolving the temperature field
with the beam and multiplying it by the size of the differential
gain. The difference is consistent with noise as a result of the noise
in the TOD. Rows 2 and 3 show maps of differential pointing. Being
spin-1, row 2 shows the error if the instrument was oriented with
$\psi_{\rm t}{=}0$ and row 3 shows the error if the telescope was
orientated with $\psi_{\rm t}{=}\pi/2$. The systematic error on a TOD
sample is simply a rotation of this spin-1 vector. The prediction is
created by using the first differential of the temperature map and
multiplying the results by half the angular size of the differential
pointing. Rows 4 and 5 are maps of the differential ellipticity. As
the differential ellipticity is a spin-2 systematic we are only able
to make the distinction between the systematic and the polarisation
using the HWP. The fields depicted in rows 4 and 5 are similar to the
Stokes $Q$ and $U$ fields respectively. The prediction was created by
using the underlying temperature field and the beam shape used in the
simulation. It was shown in \citet{2014MNRAS.442.1963W} that the
spin-2 systematic would have the form, 
\ba \Delta a^{E}_{\ell m} &=&
\sqrt{\frac{4\pi}{2\ell+1}}\Re(b^{\rm diff}_{\ell 2})a^T_{\ell
  m},\\ \label{eq:TtoE_rm} \Delta a^{B}_{\ell m} &=&
i\sqrt{\frac{4\pi}{2\ell+1}} \Im(b^{\rm diff}_{\ell 2})a^T_{\ell
  m}, \label{eq:TtoB_rm}
\ea
where $\Delta a^{E}_{\ell m}$ and $\Delta a^{B}_{\ell m}$ are the $E$
and $B$-mode of the systematic, $a^T_{\ell m}$ is the spherical
harmonic decomposition of the temperature field and $b^{\rm
  diff}_{\ell 2}$ is the second azimuthal mode of the spherical
harmonic decomposition of the difference of the two temperature beams
in the detector pair.

Figure \ref{fig:sys_maps} shows that the map-making algorithm can
correctly recover maps of the systematic errors. However, the maps are
noisy. One can imagine a situation where the noise level is too large
to see the systematic but there is a non-negligible effect on the
recovered $B$-mode power spectrum. This will be especially true when
many detectors are considered as the maps would have to be made for
each detector pair. However, since each of the systematics couple to
the temperature field, the recovered maps of the systematics will
correlate with the temperature field.

\begin{figure}
\begin{center}
\includegraphics[width=\linewidth, trim=0cm 0cm 0cm 0cm, clip=true]{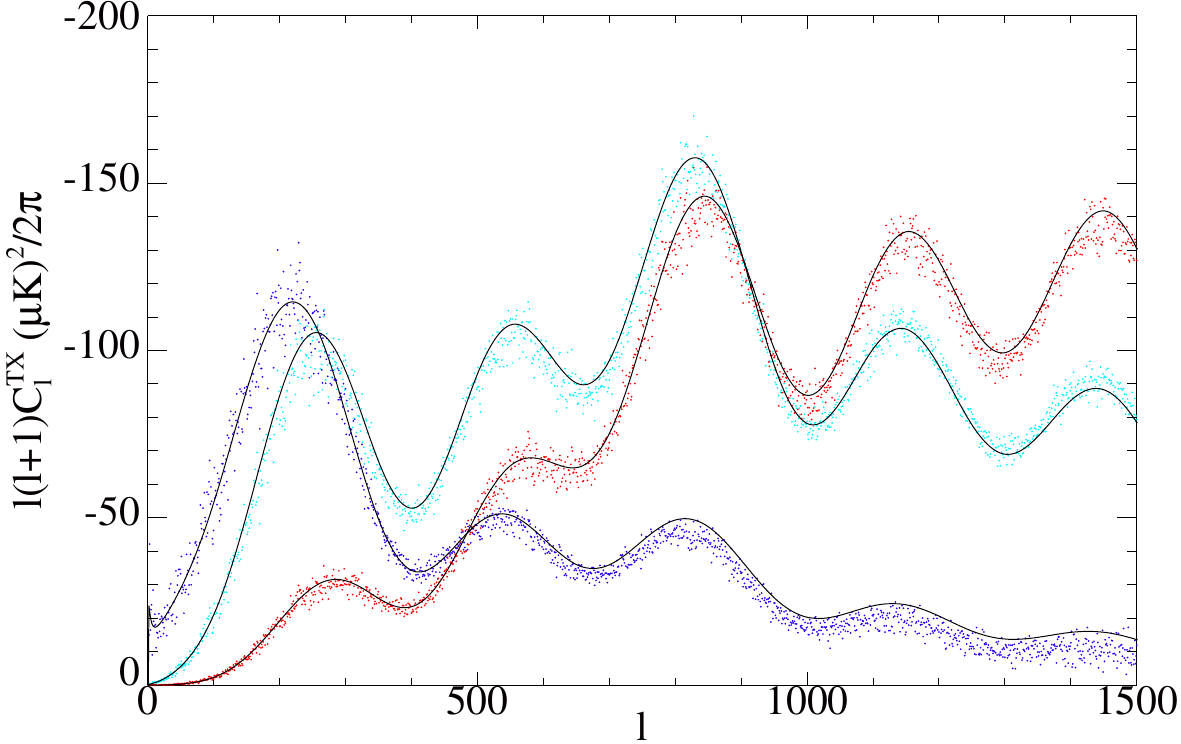} 
\caption{We plot the cross-correlation of the systematic maps (see
  Figure \ref{fig:sys_maps}) with the temperature map created using
  the TOD in simulation 3 in Section \ref{sec:sim_test}. The
  cross-power spectrum is a good method to find systematics that may
  be below the noise level in the systematic maps. The differential
  gain systematic error map cross correlated with the temperature map
  is plotted in dark blue dots. The light blue dots show the differential
  pointing maps cross correlated with the temperature map. The
  differential ellipticity map cross correlated with the temperature
  map is shown as the red dots. Over plotted are $N_0C^T_\ell$, $N_1\ell
  C^T_\ell$, $N_2\ell^2C^T_\ell$, where $N_i$ are normalisation
  factors. This demonstrates that the systematics are simply the
  temperature field convolved with the beam, or derivatives thereof.
\label{fig:sys_cor}}
\end{center}
\end{figure}

In Fig.~\ref{fig:sys_cor} we plot the cross-power spectrum between
systematic maps and the temperature map created using the TOD used in
the simulation, where we have also deconvolved for the beam. We have
over plotted $N_0C^T_\ell$, $N_1\ell C^T_\ell$, $N_2\ell^2C^T_\ell$,
where $N_i$ are normalisation factors calculated by minimising the
absolute residuals between the model and the cross-power
spectrum. This demonstrates the known result that the systematics are
simply the temperature field, or some derivative of the temperature
field. Even in a noisy systematic map the cross-power spectrum will
provide a valuable insight to the size of a systematic effect. We
recover a non-zero correlation because the systematics are present in
the TOD. If this were not the case and the TOD was clean then we would
find a cross-power spectrum consistent with zero. This provides us
with a recipe to test for the presence or absence of systematic
effects: if, in an experiment, this cross-power spectrum is shown to
be consistent with zero then the maps can be (re-)made not accounting
for this systematic. The resulting increase of noise in the map, with
respect to a binned map, would thus be kept to a minimum.

\section{Discussion} \label{sec:discuss}

We have developed two map-making algorithms to remove common
systematics that couple temperature to polarisation in differencing
CMB polarisation experiments.  The result of the map-making algorithms
is the polarisation sky smoothed with the axisymmetric part of the
beam used. The systematics we consider are differential gain, pointing
and ellipticity of two detectors in a detector pair, all of which were
shown to be an issue in the BICEP2 experiment
\citep{2014arXiv1403.4302B}. The main issue with these systematics is
the leakage from temperature to polarisation that they
create. \citet{2008PhRvD..77h3003S} showed that the coupling from
temperature to polarisation of these systematics have spin-0,1 and 2
properties respectively. We used this understanding to develop the
algorithm used here. 

The first algorithm, described in Section \ref{sec:psi_t_ext}, removes
the systematics by separating the Fourier modes of the systematics and
the Fourier mode of the polarisation, using equation
\eqref{eq:2d_for}. The technique requires a suitable scan
strategy. The angle coverage of the orientation of the telescope,
$\psi_{\rm t}$, must be extensive to allow for the different Fourier
modes to be distinguished. We have shown through simulations in
Section \ref{sec:epic_sims} that the EPIC \citep{2009arXiv0906.1188B}
scan strategy provides the required amount of angle coverage.

In Section \ref{sec:epic_sims} we demonstrated the effectiveness of the
algorithm through three simulations. Simulation 1 showed the ability of
the technique to remove differential gain and pointing when a HWP is
not used. Without a HWP the spin-2 systematic and the polarisation
signal are degenerate and, therefore, cannot be separated using this
technique. In this case we suggest using the methods proposed in
\citet{2014MNRAS.442.1963W} and \citet{2014arXiv1403.4302B}. The
technique however, can remove differential gain and pointing as they
have a different spin to the polarisation signal. In Section
\ref{sec:sim2} and \ref{sec:sim3} we presented simulations including a
HWP. In these simulations we showed that the technique can simultaneously
remove differential gain, pointing and ellipticity. {\color{black}The CMB dipole 
can contribute to the leakage from temperature to polarisation.
We demonstrated in Section \ref{sec:sim4} that the map-making
algorithm can deal with this level of leakage.}

One draw back to this technique is an increase in the statistical
noise in the resultant (cleaned) $Q$ and $U$ maps. The level of the
noise increase is dependent on the scan strategy --- the more even the
$\psi_{\rm t}$ angle coverage, the lower the increase in noise. An
ideal experiment would suffer no increase in noise. At the other
extreme where the coverage is not large enough, the matrix in equation
\eqref{eq:2d_for} becomes singular and the effective increase in noise
is infinite. Through simulations we have shown that the increase in noise
power for simulation 3 is 12\% when compared to a binned map.

The second algorithm, described in Section \ref{sec:psi_t_lim},
removes the temperature to polarisation leakage by creating a model
for total leakage as a function of the orientation of the
telescope. The combined effect of the systematic is modelled as a
smooth function of the orientation angle. With this assumption we can
then describe the combined systematic $f(x)$ by a few Legendre
polynomials. Figure \ref{fig:leg_demo} shows that a spin-2 systematic
can accurately be reconstructed by the first three Legendre
polynomials, with a $\psi_{\rm t}$ range of 0.5 rads. This $\psi_{\rm
  t}$ range was chosen to be representative of the LSPE scan strategy
\citep{2012SPIE.8446E..7AA}, where the maximum $\psi_{\rm t}$ range is
$\approx$0.5 rads. In Section \ref{sec:balloon}, we demonstrated that
the algorithm can remove the temperature to polarisation leakage from
differential gain, pointing and ellipticity in a simulation where the
LSPE scan strategy was used with a stepped HWP. As with the
extensive $\psi_{\rm t}$ range technique there is an increase in the
statistical noise of the polarisation maps when using this
technique. Through simulations we showed that the increase of noise
using the LSPE scan strategy was 12\%.

In Section \ref{sec:identify_sys} we have presented a method to
identify if systematics are present in a TOD. The extensive $\psi_{\rm
  t}$ range algorithm separates the systematics into Fourier modes and
generates an estimate of the polarisation free of these
systematics. It also at the same time creates an estimate of the
systematic. We showed in Fig.~\ref{fig:sys_maps} the maps of the
systematics recovered from simulation 3 (see Section
\ref{sec:sim3}). These maps can be used to identify if a systematic is
present. However, the noise in the TOD and the relative size of the
systematic effect could render the reconstructed maps too noisy to see
the systematic easily. To increase the signal-to-noise, we suggest
calculating the cross-power spectrum of these systematic maps with the
temperature map. As the systematics we are considering are due to
temperature to polarisation leakage then, if the systematic is
present, we expect to see a non-zero cross-correlation. We showed the
cross-power spectrum between the systematic maps and the temperature
map created using the TOD in Fig.~\ref{fig:sys_cor}. This technique
can be used to identify if systematics are present. This is crucial to
test the validity of an experiment's polarisation maps, but could also
be used to identify whether a systematic must be removed using this
technique. Accounting only for those systematics that are actually
present in the TOD would minimise the increase in noise associated with
the correction algorithms developed in this work.

\section*{Acknowledgments}

CGRW acknowledges the award of a STFC quota studentship. AB and MLB are
grateful to the European Research Council for support through the
award of an ERC Starting Independent Researcher Grant (EC FP7 grant
number 280127). MLB also thanks the STFC for the award of Advanced and
Halliday fellowships (grant number ST/I005129/1). The authors thank 
Giampaolo Pisano and Luca Lamagna for providing the simulated LSPE
beam used in Section \ref{sec:balloon} plotted in Figure 1 of 
\citet{2014MNRAS.442.1963W}.
Some of the results in this paper have been derived using
the {\sevensize HEALPix}~\citep{2005ApJ...622..759G} package.
{\color{black} We thank the referee for their useful comments.}
 
\bibliographystyle{mnras}
\bibliography{references}



\label{lastpage}

\end{document}